# Shallow Embedding of Type Theory is Morally Correct*

Ambrus Kaposi[0000−0001−9897−8936], András Kovács[0000−0002−6375−9781], and Nicolai Kraus[0000−0002−8729−4077]

Eötvös Loránd University

**Abstract.** There are multiple ways to formalise the metatheory of type theory. For some purposes, it is enough to consider specific *models* of a type theory, but sometimes it is necessary to refer to the *syntax*, for example in proofs of canonicity and normalisation. One option is to embed the syntax deeply, by using inductive definitions in a proof assistant. However, in this case the handling of definitional equalities becomes technically challenging. Alternatively, we can reuse conversion checking in the metatheory by *shallowly embedding* the object theory. In this paper, we consider the *standard model* of a type theoretic object theory in Agda. This model has the property that all of its equalities hold definitionally, and we can use it as a shallow embedding by building expressions from the components of this model. However, if we are to reason soundly about the syntax with this setup, we must ensure that distinguishable syntactic constructs do not become provably equal when shallowly embedded. First, we prove that shallow embedding is injective up to definitional equality, by modelling the embedding as a syntactic translation targeting the metatheory. Second, we use an implementation hiding trick to disallow illegal propositional equality proofs and constructions which do not come from the syntax. We showcase our technique with very short formalisations of canonicity and parametricity for Martin-Löf type theory. Our technique only requires features which are available in all major proof assistants based on dependent type theory.

## 1 Introduction

Martin-Löf type theory [32] (MLTT) is a formal system which can be used for writing and verifying programs, and also for formalising mathematics. Proof assistants and dependently typed programming languages such as Agda [43], Coq [33], Idris [9], and Lean [36] are based on MLTT and its variations.

* The research has been supported by the European Union, co-financed by the European Social Fund (EFOP-3.6.2-16-2017-00013, Thematic Fundamental Research Collaborations Grounding Innovation in Informatics and Infocommunications and FOP-3.6.3-VEKOP-16-2017-00002) and COST Action EUTypes CA15123.

Specific versions of MLTT have many interesting properties, such as *canonicity*, *normalisation* or *parametricity*. Normalisation in particular is practically significant, since it enables decidable conversion checking and thus decidable type checking. These properties are of metatheoretic nature; in other words, they are answers to questions *about* type theory, rather than questions *inside* type theory. We wish to effectively study these questions in a formal and machine-checked setting.

### 1.1 Technical Challenges of Deep Embeddings

We refer to the type theory that we wish to study as the *object (type) theory*. If we want to use Agda (or another proof assistant) to study it, the most direct way is to use native inductive definitions to represent the syntax. This is called a *deep embedding*. Such an embedding could be an inductive type representing syntactic expressions Expr, with a constructor for every kind of term former. Examples for such constructors are the following:

$$\begin{aligned}
\mathsf{Pi} &: \mathsf{Expr} \to \mathsf{Expr} \to \mathsf{Expr} \\
\mathsf{lam} &: \mathsf{Expr} \to \mathsf{Expr} \\
\mathsf{app} &: \mathsf{Expr} \to \mathsf{Expr} \to \mathsf{Expr}
\end{aligned}$$

The idea is simple: Pi takes two expressions $e_1, e_2$ as arguments, and if these represent a type $A$ and a type family $B$ over $A$, then $\mathsf{Pi}\, e_1\, e_2$ represents the corresponding Π-type. Similarly, lam represents λ-abstraction and app application.

Of course, this inductive definition of Expr does not ensure that every expression "makes sense"; e.g. $\mathsf{Pi}\, e_1\, e_2$ will not make sense unless $e_1$ and $e_2$ are of the form described above. We need to additionally define inductive relations which express well-formedness and typing for specific syntactic constructs. This way of defining raw terms together with well-formedness relations is called an *extrinsic* approach.

Depending on the available notion of inductive types in the metatheory, we can use more abstract representations. For example, if inductive-inductive types [37] are available, then we can define a syntax which contains only well-formed terms [10]. In this case, we have an *intrinsic* definition for the syntax. We have the following signature for the type constructors of the embedded syntax, respectively for contexts, types, substitutions and terms:

$$\begin{aligned}
\mathsf{Con} &: \mathsf{Set} \\
\mathsf{Ty} &: \mathsf{Con} \to \mathsf{Set} \\
\mathsf{Sub} &: \mathsf{Con} \to \mathsf{Con} \to \mathsf{Set} \\
\mathsf{Tm} &: (\Gamma : \mathsf{Con}) \to \mathsf{Ty}\,\Gamma \to \mathsf{Set}
\end{aligned}$$

However, with the intrinsic inductive-inductive definitions we also need separate inductive relations expressing definitional equality. We can avoid these relations by using a *quotient inductive* [29,2] syntax instead. This way, definitional equality is given by *equality constructors*. For example, associativity of type substitution would be given as the following $[\circ]$ equality, where we also introduce substitution composition and type substitution first, and implicitly quantify over variables:

$$
\begin{aligned}
\cdot \circ \cdot &: \mathsf{Sub}\,\Theta\,\Delta \to \mathsf{Sub}\,\Gamma\,\Theta \to \mathsf{Sub}\,\Gamma\,\Delta \\
\cdot[\cdot] &: \mathsf{Ty}\,\Delta \to \mathsf{Sub}\,\Gamma\,\Delta \to \mathsf{Ty}\,\Gamma \\
[\circ] &: (A\,[\sigma])\,[\delta] = A\,[\sigma \circ \delta]
\end{aligned}
$$

The quotient inductive definition allows higher-level reasoning than the purely inductive-inductive one. In the former case, every metatheoretic construction automatically respects definitional equality in the syntax, since it is identified with meta-level propositional equality. In the latter case, object-level definitional equality is just a relation, and we need to explicitly prove preservation in many cases.

However, even with quotient induction, there are major technical challenges in formalising metatheory, and an especially painful issue is the obligation to explicitly refer to conversion rules even in very simple constructions. For example, we might want to take the zeroth de Bruijn index with type $\mathsf{Bool}$ in some extended $\Gamma \blacktriangleright \mathsf{Bool}$ typing context. For this, we first need a weakening substitution declared in the syntax (or admissible from the syntax):

$$\mathsf{weaken} : \mathsf{Sub}\,(\Gamma \blacktriangleright A)\,\Gamma$$

Now, we are able to give a general type for the zeroth de Bruijn index:

$$\mathsf{vzero} : \mathsf{Tm}\,(\Gamma \blacktriangleright A)\,(A[\mathsf{weaken}])$$

The weakening is necessary because $A$ has type $\mathsf{Ty}\,\Gamma$, but we also want to mention it in the $\Gamma \blacktriangleright A$ context.

Now, we might try to use $\mathsf{vzero}$ to get a term with type $\mathsf{Tm}\,(\Gamma \blacktriangleright \mathsf{Bool})\,\mathsf{Bool}$. However, we only get $\mathsf{vzero} : \mathsf{Tm}\,(\Gamma \blacktriangleright \mathsf{Bool})\,(\mathsf{Bool}[\mathsf{weaken}])$. We also need to refer to the computation rule for substituting $\mathsf{Bool}$ which just forgets about the substitution:

$$\mathsf{Bool}[] : \mathsf{Bool}[\sigma] = \mathsf{Bool}$$

Hence, the desired term needs to involve transporting over the $\mathsf{Bool}[]$ equation:

$$
\begin{aligned}
&\mathsf{vzeroBool} : \mathsf{Tm}\,(\Gamma \blacktriangleright \mathsf{Bool})\,\mathsf{Bool} \\
&\mathsf{vzeroBool} :\equiv \mathsf{transport}_{(\mathsf{Tm}\,\Gamma)}\,\mathsf{Bool}[]\,\mathsf{vzero}
\end{aligned}
$$

This phenomenon arises with extrinsic and purely inductive-inductive syntaxes as well; in those cases, instead of transporting along an equation, we need to invoke a conversion rule for term typing. For extrinsic syntaxes, we additionally have a choice between implicit and explicit substitution, but this choice does not change the picture either.

Hence, all of the mentioned deeply embedded syntaxes require constructing explicit derivations of definitional equalities. In more complex examples, this is a technical burden which is often humanly impossible to handle. Also, proof assistants are often unable to check formalisations within sensible time because of the huge size of the involved proof terms.

### 1.2 Reflecting Definitional Equality

To eliminate explicit derivations of conversion, the most promising approach is to reflect object-level definitional equality as meta-level definitional equality. If this is achieved, then all conversion derivations can be essentially replaced by proofs of reflexivity, and the meta-level typechecker would implicitly construct all derivations for us.

How can we achieve this? We might consider extensional type theory with general equality reflection, or proof assistants with limited equality reflection. In Agda there is support for the latter using rewrite rules [12], which we have examined in detail for the previously described purposes. In Agda, we can just postulate the syntax of the object theory, and try to reflect the equations. This approach does work to some extent, but there are significant limitations:

- Type-directed equalities cannot be reflected, such as $\eta$-rules for empty substitutions and unit types, or definitional proof irrelevance for propositions. Rewrite rules must be syntax-directed and have a fixed direction of rewriting.
- Rewrite rules yield poor evaluation performance and hence poor type checking performance, because they are implemented using a general mechanism which does not know anything about the domain, unlike the meta-level conversion checker.
- In the current Agda implementation (version 2.6), rewrite rules are not flexible enough to capture all desired computational behavior. For example, the left hand side of a rewrite rule is treated as a rigid expression which is not refined during the matching of the rule. Given an $f : \mathsf{Bool} \to \mathsf{Bool} \to \mathsf{Bool}$ function, if we add the rewrite rule $\forall x. f\, x\, (\mathsf{not}\, x) = \mathsf{true}$, the expession $f\, \mathsf{true}\, \mathsf{false}$ will not be rewritten to $\mathsf{true}$, since it does not rigidly match the $\mathsf{not}\, x$ on the left hand side. In practice, this means that an unbounded number of special-cased rules are required to reflect equalities for a type theory. Lifting all the restricting assumptions in the implementation of rewrite rules would require non-trivial research effort.

It seems to be difficult to capture the equational theory of a dependent object theory with general-purpose implementations of equality reflection. In the future, robust equality reflection for conversion rules may become available, but until then we have to devise workarounds. If the object theory is similar enough to the metatheory, we can reuse meta-level conversion checking using a *shallow embedding*.

In this paper we describe such a shallow embedding. The idea is that in the *standard model* of the object theory equations already hold definitionally, and so it would be convenient to reason about expressions built from the standard model as if they came from arbitrary models, e.g. from the syntax.

However, we should only use shallow embeddings in morally correct ways: only those equations should hold in the shallow embedding that also hold in the deeply embedded syntax. In other words, we should be able in principle to translate every formalisation which uses shallow embedding to a formalisation which uses deeply embedded syntax.

To address this, first we prove that shallow embedding is injective up to *definitional equality*: the metatheory can only believe two embedded terms definitionally equal if they are already equal in the object theory. This requires us to look at both the object theory and the metatheory from an external point of view and reason about embedded meta-level terms as pieces of syntax.

Second, we describe a method for hiding implementation details of the standard model, which prevents constructing terms which do not have syntactic counterparts and which also disallows morally incorrect *propositional equalities*. This hiding is realised with import mechanisms; we do not formally model it, but it is reasonable to believe that it achieves the intended purposes.

### 1.3 Contributions

*In order to reason about the metatheory of type theory in a proof assistant, we present a version of shallow embedding which combines the advantage of shallow embeddings (many definitional equalities) with the advantage of deep embeddings (no unjustified equalities).*

In detail:

1. We formalise in Agda the standard "Set" model (metacircular interpretation [22]) of a variant of MLTT with a predicative universe hierarchy, Π-types, Booleans, Σ-types and identity types (Section 3). All equalities hold definitionally in this model. A variation of this (see below) is the model we propose for metatheoretic reasoning.
2. For an arbitrary model of the object theory, we construct the *termified* model (Section 4), where contexts, types, substitutions and terms are all modelled by closed terms. We formalise the shallow embedding into Agda as the interpretation of the object syntax into its termified model. We prove that this translation is injective (Section 5), thereby showing that definitional equality of shallowly embedded terms coincides with object-theoretic definitional equality. This result holds *externally* to Agda (like parametricity): we need to step one level up and consider the syntax of Agda as well. Additionally, we show that internally to Agda, injectivity of the standard interpretation is not provable.
3. We describe a way of hiding the implementation of the standard model (Section 6), in order to rule out constructions and equality proofs which are not available in the object syntax.
4. Using shallowly embedded syntax, we provide a concise formalisation of canonicity for MLTT (Section 7.2), using a proof-relevant logical predicate model in a manner similar to [14] and [27]. We also provide a formalisation of a syntactic parametricity translation [6] of MLTT in Section 7.1.

The contents of Sections 3, 4, 6 and 7 were formalised [30] in Agda. Additional documentation about technicalities is provided alongside the formalisation.

### 1.4 Related Work

Work on embedding the syntax of type theory in type theory spans a whole spectrum from fully deep embeddings through partly deep embeddings to fully shallow ones.

Deep embeddings give maximal flexibility but at the high price of explicit handling of definitional equality derivations. Extrinsic deep embeddings of type theory are given in Agda [18,1] and Coq [44]. Meta Coq provides an extrinsic deep embedding of the syntax of almost all of Coq inside Coq [5]. An intrinsic deep embedding with explicit conversion relations using inductive-inductive types is given in [10] and another one using inductive-recursive types is described by [16].

Quotient inductive-inductive types are used in [3,4] to formalise type theory in a bit more shallow way reusing propositional equality of the metatheory to represent conversion of the object theory.

Higher-order abstract syntax (HOAS) [39,23] uses shallow embedding for the substitution calculus part of the syntax while the rest (e.g. term formers such as $\lambda$ and application) are given deeply, using the function space of the metalanguage to represent binders. It has been used to embed simpler languages in type theory [17,40,11], however, to our knowledge, not type theory itself.

McBride [34] uses a mixture of deep and shallow embeddings to embed an intrinsic syntax of type theory into Agda. In this work, inductively defined types and terms are given mutually with their standard interpretation, and while there are deep term and type codes, all *indexing* in the syntax is over the standard model. In a sense, this is an extension of inductive-recursive type codes to codes of terms as well. This gives a usability improvement compared to deep embedding as equality of indices is decided by the metatheory. However, definitional equality of terms still has to be represented deeply.

Shallow embedding has been used to formalise constructions on the syntax of type theory. [26,8,42] formalise the correctness of syntactic translations using shallow embeddings in Coq. [28,29] formalise syntactic translations and models of type theory depending on previous shallow models. Our work provides a framework in which these previous formalisations could be rewritten in a more principled way.

Reflection provides an interface between shallow and deep embeddings. Meta Coq [5] provides a mechanism to reify shallow Coq terms as deeply embedded syntax. The formalisation happens shallowly, making use of the typechecker of Coq, and deeply embedded terms are obtained after reification. The motivation is very similar to ours, but their syntax is extrinsic while we use an intrinsic syntax.

More generally, using type theory as an internal language of a model can be seen as working in a shallow embedding. Synthethic homotopy theory (e.g. [24]) can be seen as a shallow embedding in type theory, compared to a deep embedding where homotopy theory is built up from the ground analytically. [38]

uses MLTT extended with some axioms to formalise arguments about a presheaf model, [15] uses MLTT as the internal language of a cubical set model, [29] uses MLTT as the internal language of a categories-with-families model.

Our wrapped shallow embedding (Section 6) resembles the method by Dan Licata [31] to add higher inductive types to Agda with eliminators computing definitionally on point constructors. He also uses an implementation hiding to disallow pattern matching but retain definitional behaviour.

## 2 The Involved Theories

In this paper, we altogether need to involve three different theories. We give a quick overview below, then describe them and the used notation in this section.

1. **Agda**, which we use in two ways: as a metatheory when using shallow embedding, but also as an object theory, when we study embedding from an external point of view. In the latter case, we only talk about a small subset of Agda's syntax which is relevant to the current paper.
2. The **external metatheory**. We assume that this is a conventional extensional type theory with a universe hierarchy. However, we are largely agnostic and set theory with a suitable notion of universe hierarchy would be adequate as well. We primarily use the external metatheory to reason about Agda's syntax. However, since this metatheory is extensional, we can omit all coercions and transports when working inside it informally, and thus we also use it to obtain a readable notation.
3. The **object theory**, which we wish study by shallow embedding into Agda. We single out a particular version of MLTT as object theory, and describe it in detail. However, our shallow embedding should work for a wider range of object theories; we expand on this in Section 8.1.

### 2.1 Agda

Agda is a proof assistant based on intensional type theory. When we present definitions in Agda, we use a `monospace` font. We describe below the used features and notation.

Universes are named `Set i`. Also, we use universe polymorphism which allows us to quantify over `(i : Level)`. We use `zero` and `suc` for the zero and successor levels, and `i ⊔ j` for taking least upper bounds of levels.

Dependent functions are notated `(x : A) → B`. There is also an implicit function space `{x : A} → B`, such that any expression with this type is implicitly applied to an inferred `A` argument. In this paper, we also use implicit quantification over variables in type signatures. For example, instead of declaring a type as `f : {A : Set} → A → A`, we may write `f : A → A`. This shorthand (although supported in the latest 2.6 version of Agda) is not used in the actual formalisations.

We also use Σ types, unit types, Booleans and propositional equality. There are some names which coincide in the object theory and in agda, and we disambiguate them with a `m.` prefix (which stands for "meta"). So, we use `m.Σ A B` for dependent pairs with `(t m., u)` as constructor and `m.fst` and `m.snd` as projections. We use `m.Bool`, `m.true` and `m.false` for Booleans. We use `m.⊤` for the unit type with constructor `m.tt`, and use `t ≡ u` for propositional equality with `m.refl` and `m.J`.

## 2.2 The External Metatheory

This is an extensional type theory, with predicative universes $\mathsf{Set}_i$, dependent functions $(x : A) \to B$, and dependent pairs as $(x : A) \times B$. Propositional equality is denoted $\cdot = \cdot$, with constructor $\mathsf{refl}$. We have equality reflection, which means that if $p : t = u$ is derivable, then $t$ and $u$ are definitionally equal. We also have uniqueness of identity proofs, meaning that for any $p, q : t = u$ we also have $p = q$.

## 2.3 The Object Type Theory

We take an algebraic approach to the syntax and models of type theory. There is an *algebraic signature* for the object type theory, which can be viewed as a large record type, listing all syntactic constructions along with the equations for definitional equality. *Models* of a type theory are particular inhabitants of this large record type, and the *syntax* of a type theory is the *initial* model in the category of models, where morphisms are given by structure-preserving families of functions. The setup can be compared to groups, a more familiar algebraic structure: there is a signature for groups, models are particular groups, morphisms are group homomorphisms, and the initial group ("syntax") is the trivial group (free group over the empty set). A *displayed model* over a model $\mathcal{M}$ is a way of encoding a model together with a morphism into $\mathcal{M}$. Displayed models can be viewed as containing induction motives and methods for a theory (following the nomenclature of [35]), hence we need this notion to talk about induction over the syntax. For instance, a displayed model for the theory of natural numbers contains a family $P : \mathbb{N} \to \mathsf{Set}$ (the induction motive) together with induction methods showing that $P$ is inhabited at $\mathsf{zero}$ and taking successors preserves $P$. A generic method for deriving the notions of model, morphism and displayed model from a signature is given in [29].

More concretely, our object type theory is given in Figures 1a and 1b as a category with families (CwF) [20] extended with additional type formers. We present the signature of the object theory in an extensional notation, which allows us to omit transports along equations. We also implicitly quantify over variables occurring in types, and leave these parameters implicit when we apply functions as well. Additionally, we extend the usual notion of CwF with indexing by metatheoretic natural numbers, which stand for universe levels.

This notion of model yields a syntax with explicit substitutions. The core structural rules and the theory of substitutions are described by the components

$$
\begin{array}{ll}
\mathsf{Con} & : \mathbb{N} \to \mathsf{Set} \\
\mathsf{Ty} & : \mathbb{N} \to \mathsf{Con}\, i \to \mathsf{Set} \\
\mathsf{Sub} & : \mathsf{Con}\, i \to \mathsf{Con}\, j \to \mathsf{Set} \\
\mathsf{Tm} & : (\Gamma : \mathsf{Con}\, i) \to \mathsf{Ty}\, j\, \Gamma \to \mathsf{Set} \\
\mathsf{id} & : \mathsf{Sub}\, \Gamma\, \Gamma \\
\cdot \circ \cdot & : \mathsf{Sub}\, \Theta\, \Delta \to \mathsf{Sub}\, \Gamma\, \Theta \to \mathsf{Sub}\, \Gamma\, \Delta \\
\mathsf{ass} & : (\sigma \circ \delta) \circ \nu = \sigma \circ (\delta \circ \nu) \\
\mathsf{idl} & : \mathsf{id} \circ \sigma = \sigma \\
\mathsf{idr} & : \sigma \circ \mathsf{id} = \sigma \\
\cdot[\cdot] & : \mathsf{Ty}\, i\, \Delta \to \mathsf{Sub}\, \Gamma\, \Delta \to \mathsf{Ty}\, i\, \Gamma \\
\cdot[\cdot] & : \mathsf{Tm}\, \Delta\, A \to (\sigma : \mathsf{Sub}\, \Gamma\, \Delta) \to \mathsf{Tm}\, \Gamma\, (A[\sigma]) \\
[\mathsf{id}] & : A[\mathsf{id}] = A \\
[\circ] & : A[\sigma \circ \delta] = A[\sigma][\delta] \\
[\mathsf{id}] & : t[\mathsf{id}] = t \\
[\circ] & : t[\sigma \circ \delta] = t[\sigma][\delta] \\
\bullet & : \mathsf{Con}\, 0 \\
\epsilon & : \mathsf{Sub}\, \Gamma\, \bullet \\
\bullet\eta & : (\sigma : \mathsf{Sub}\, \Gamma\, \bullet) = \epsilon \\
\cdot \triangleright \cdot & : (\Gamma : \mathsf{Con}\, i) \to \mathsf{Ty}\, j\, \Gamma \to \mathsf{Con}\, (i \sqcup j) \\
\cdot , \cdot & : (\sigma : \mathsf{Sub}\, \Gamma\, \Delta) \to \mathsf{Tm}\, \Gamma\, (A[\sigma]) \to \mathsf{Sub}\, \Gamma\, (\Delta \triangleright A) \\
\mathsf{p} & : \mathsf{Sub}\, (\Gamma \triangleright A)\, \Gamma \\
\mathsf{q} & : \mathsf{Tm}\, (\Gamma \triangleright A)\, (A[\mathsf{p}]) \\
\triangleright\beta_1 & : \mathsf{p} \circ (\sigma, t) = \sigma \\
\triangleright\beta_2 & : \mathsf{q}[\sigma, t] = t \\
\triangleright\eta & : (\mathsf{p}, \mathsf{q}) = \mathsf{id} \\
,\circ & : (\sigma, t) \circ \nu = (\sigma \circ \nu, t[\nu]) \\
\Pi & : (A : \mathsf{Ty}\, i\, \Gamma) \to \mathsf{Ty}\, j\, (\Gamma \triangleright A) \to \mathsf{Ty}\, (i \sqcup j)\, \Gamma \\
\mathsf{lam} & : \mathsf{Tm}\, (\Gamma \triangleright A)\, B \to \mathsf{Tm}\, \Gamma\, (\Pi\, A\, B) \\
\mathsf{app} & : \mathsf{Tm}\, \Gamma\, (\Pi\, A\, B) \to \mathsf{Tm}\, (\Gamma \triangleright A)\, B \\
\Pi\beta & : \mathsf{app}\, (\mathsf{lam}\, t) = t \\
\Pi\eta & : \mathsf{lam}\, (\mathsf{app}\, t) = t \\
\Pi[] & : (\Pi\, A\, B)[\sigma] = \Pi\, (A[\sigma])\, (B[\sigma^{\uparrow}]) \\
\mathsf{lam}[] & : (\mathsf{lam}\, t)[\sigma] = \mathsf{lam}\, (t[\sigma^{\uparrow}]) \\
\Sigma & : (A : \mathsf{Ty}\, i\, \Gamma) \to \mathsf{Ty}\, j\, (\Gamma \triangleright A) \to \mathsf{Ty}\, (i \sqcup j)\, \Gamma \\
\cdot , \cdot & : (u : \mathsf{Tm}\, \Gamma\, A) \to \mathsf{Tm}\, \Gamma\, (B[\mathsf{id}, u]) \to \mathsf{Tm}\, \Gamma\, (\Sigma\, A\, B) \\
\mathsf{fst} & : \mathsf{Tm}\, \Gamma\, (\Sigma\, A\, B) \to \mathsf{Tm}\, \Gamma\, A \\
\mathsf{snd} & : (t : \mathsf{Tm}\, \Gamma\, (\Sigma\, A\, B)) \to \mathsf{Tm}\, \Gamma\, (B[\mathsf{id}, \mathsf{fst}\, t])
\end{array}
$$

**Fig. 1a.** The object type theory as a generalised algebraic structure. $\sigma^{\uparrow}$ abbreviates $(\sigma \circ \mathsf{p}, \mathsf{q})$.

$$
\begin{array}{ll}
\Sigma\beta_1 & : \mathsf{fst}\,(u,v) = u \\
\Sigma\beta_2 & : \mathsf{snd}\,(u,v) = v \\
\Sigma\eta & : (\mathsf{fst}\,t, \mathsf{snd}\,t) = t \\
\Sigma[] & : (\Sigma\,A\,B)[\sigma] = \Sigma\,(A[\sigma])\,(B[\sigma^{\uparrow}]) \\
,[] & : (u,v)[\sigma] = (u[\sigma], v[\sigma]) \\
\top & : \mathsf{Ty}\,0\,\Gamma \\
\mathsf{tt} & : \mathsf{Tm}\,\Gamma\,\top \\
\top\eta & : (t : \mathsf{Tm}\,\Gamma\,\top) = \mathsf{tt} \\
\top[] & : \top[\sigma] = \top \\
\mathsf{tt}[] & : \mathsf{tt}[\sigma] = \mathsf{tt} \\
\mathsf{U} & : (i : \mathbb{N}) \to \mathsf{Ty}\,(i+1)\,\Gamma \\
\underline{\cdot} & : \mathsf{Tm}\,\Gamma\,(\mathsf{U}\,i) \to \mathsf{Ty}\,i\,\Gamma \\
\mathsf{c} & : \mathsf{Ty}\,i\,\Gamma \to \mathsf{Tm}\,\Gamma\,(\mathsf{U}\,i) \\
\mathsf{U}\beta & : \underline{\mathsf{c}\,A} = A \\
\mathsf{U}\eta & : \mathsf{c}\,\underline{a} = a \\
\mathsf{U}[] & : (\mathsf{U}\,i)[\sigma] = (\mathsf{U}\,i) \\
\underline{[]} & : \underline{a}[\sigma] = \underline{a[\sigma]} \\
\mathsf{Bool} & : \mathsf{Ty}\,0\,\Gamma \\
\mathsf{true} & : \mathsf{Tm}\,\Gamma\,\mathsf{Bool} \\
\mathsf{false} & : \mathsf{Tm}\,\Gamma\,\mathsf{Bool} \\
\mathsf{if} & : (C : \mathsf{Ty}\,i\,(\Gamma \triangleright \mathsf{Bool})) \to \mathsf{Tm}\,\Gamma\,(C[\mathsf{id},\mathsf{true}]) \to \mathsf{Tm}\,\Gamma\,(C[\mathsf{id},\mathsf{false}]) \to \\
 & \quad (t : \mathsf{Tm}\,\Gamma\,\mathsf{Bool}) \to \mathsf{Tm}\,\Gamma\,(C[\mathsf{id},t]) \\
\mathsf{Bool}\beta_1 & : \mathsf{if}\,C\,u\,v\,\mathsf{true} = u \\
\mathsf{Bool}\beta_2 & : \mathsf{if}\,C\,u\,v\,\mathsf{false} = v \\
\mathsf{Bool}[] & : \mathsf{Bool}[\sigma] = \mathsf{Bool} \\
\mathsf{true}[] & : \mathsf{true}[\sigma] = \mathsf{true} \\
\mathsf{false}[] & : \mathsf{false}[\sigma] = \mathsf{false} \\
\mathsf{if}[] & : (\mathsf{if}\,C\,u\,v\,t)[\sigma] = \mathsf{if}\,(C[\sigma^{\uparrow}])\,(u[\sigma])\,(v[\sigma])\,(t[\sigma]) \\
\mathsf{Id} & : (A : \mathsf{Ty}\,i\,\Gamma) \to \mathsf{Tm}\,\Gamma\,A \to \mathsf{Tm}\,\Gamma\,A \to \mathsf{Ty}\,i\,\Gamma \\
\mathsf{refl} & : (u : \mathsf{Tm}\,\Gamma\,A) \to \mathsf{Tm}\,\Gamma\,(\mathsf{Id}\,A\,u\,u) \\
\mathsf{J} & : \big(C : \mathsf{Ty}\,i\,(\Gamma \triangleright A \triangleright \mathsf{Id}\,(A[\mathsf{p}])\,(u[\mathsf{p}])\,0)\big) \to \mathsf{Tm}\,\Gamma\,(C[\mathsf{id},u,\mathsf{refl}\,u]) \to \\
 & \quad (e : \mathsf{Tm}\,\Gamma\,(\mathsf{Id}\,A\,u\,v)) \to \mathsf{Tm}\,\Gamma\,(C[\mathsf{id},v,e[\mathsf{p}]]) \\
\mathsf{Id}\beta & : \mathsf{J}\,C\,w\,(\mathsf{refl}\,u) = w \\
\mathsf{Id}[] & : (\mathsf{Id}\,A\,u\,v)[\sigma] = \mathsf{Id}\,(A[\sigma])\,(u[\sigma])\,(v[\sigma]) \\
\mathsf{refl}[] & : (\mathsf{refl}\,u)[\sigma] = \mathsf{refl}\,(u[\sigma]) \\
\mathsf{J}[] & : (\mathsf{J}\,C\,w\,e)[\sigma] = \mathsf{J}\,(C[\sigma^{\uparrow\uparrow}])\,(w[\sigma])\,(e[\sigma])
\end{array}
$$

**Fig. 1b.** The object type theory as a generalised algebraic structure. $\sigma^{\uparrow}$ abbreviates $(\sigma \circ \mathsf{p}, \mathsf{q})$.

from $\mathsf{Con}$ to $,\circ$. Contexts ($\mathsf{Con}$) and substitutions ($\mathsf{Sub}$) form a category ($\mathsf{id}$ to $\mathsf{idr}$). There is a contravariant, functorial action of substitutions on types and terms ($\cdot[\cdot]$ to $[\circ]$), thus types (of fixed level) form a presheaf on the category of contexts and terms form a presheaf on the category of elements of this presheaf. The empty context ($\bullet$) is the terminal object.

Contexts can be extended by $\cdot \triangleright \cdot$. Substitutions can be viewed as abstract lists of terms, with $\cdot, \cdot$ allowing us to extend a substitution with a term. We can also take the "tail" and the "head" of an extended $\sigma : \mathsf{Sub}\,\Gamma\,(\Delta \triangleright A)$ substitution; the tail is given by $\mathsf{p} \circ \sigma : \mathsf{Sub}\,\Gamma\,\Delta$, and the head is given by $\mathsf{q}[\sigma] : \mathsf{Tm}\,\Gamma\,A[\mathsf{p}]$. $\mathsf{p}$ is usually called a weakening substitution, and $\mathsf{q}$ corresponds to the zeroth de Bruijn index. We denote $n$-fold composition of the weakening substitution $\mathsf{p}$ by $\mathsf{p}^n$ (where $\mathsf{p}^0 = \mathsf{id}$), and we denote De Bruijn indices the following way: $\mathsf{v}^0 := \mathsf{q}$, $\mathsf{v}^1 := \mathsf{q}[\mathsf{p}]$, $\ldots$, $\mathsf{v}^n := \mathsf{q}[\mathsf{p}^n]$. We define lifting of a substitution $\sigma : \mathsf{Sub}\,\Gamma\,\Delta$ by $\sigma^{\uparrow} : \mathsf{Sub}\,(\Gamma \triangleright A[\sigma])\,(\Delta \triangleright A) := (\sigma \circ \mathsf{p}, \mathsf{q})$. We observe that it has the property ${}^{\uparrow}[] : (\sigma^{\uparrow}) \circ (\delta, t) = (\sigma \circ \delta, t)$.

$\Pi$-types are characterised by a natural isomorphism between $\mathsf{Tm}\,\Gamma\,(\Pi\,A\,B)$ and $\mathsf{Tm}\,(\Gamma \triangleright A)\,B$, with $\mathsf{lam}$ and $\mathsf{app}$ being the morphism components. This notion of application is different from the conventional one, but in our setting with explicit substitutions, the two applications are inter-derivable, and our $\mathsf{app}$ is simpler to interpret in models. We define conventional application as $t\,\$\,u := (\mathsf{app}\,t)[\mathsf{id}, u]$. $A \Rightarrow B$ abbreviates non-dependent functions, and is defined as $\Pi\,A\,(B[\mathsf{p}])$.

$\Sigma$-types are given by the constructor $\cdot, \cdot$ and projections $\mathsf{fst}$ and $\mathsf{snd}$, and we also support the $\eta$-law. There is a unit type $\top$ with one constructor $\mathsf{tt}$ and an $\eta$-law. We have a hierarchy of universes, given by natural isomorphisms between $\mathsf{Ty}\,i\,\Gamma$ and $\mathsf{Tm}\,\Gamma\,(\mathsf{U}\,i)$ for every $i$. The isomorphism consists of a coding morphism ($\mathsf{c}$) and a decoding morphism, denoted by underlining $\underline{\cdot}$. This presentation of universes is due to Thierry Coquand, and has been used before in [25] for instance. In the Agda formalisations, where we cannot underline, we write `El` for the decoding morphism.

We also have a propositional identity type $\mathsf{Id}$, with usual constructor $\mathsf{refl}$ and elimination $\mathsf{J}$ with definitional $\beta$-rule.

Note that terms of $\Pi$-, $\Sigma$- and $\mathsf{U}$-types are all characterized by natural isomorphisms, with substitution laws corresponding to naturality conditions. Hence, we only need to state naturality in one direction, and the other direction can be derived. For example, we only state the $\underline{[]}$ substitution rule, and the other law for substituting $\mathsf{c}$ can be derived.

*Remark.* It is important that we present the notion of signature in extensional type theory instead of in Agda. The reason is that many components in the signature are well-typed only up to previous equations in the signature, and hence would need to include transports in intensional settings. The simplest example for this is the $\triangleright\beta_2$ component with type $\mathsf{q}[\sigma, t] = t$. The left side of the equation has type $\mathsf{Tm}\,\Gamma\,(A[\mathsf{p}][\sigma, t])$, while the right side has type $\mathsf{Tm}\,\Gamma\,(A[\sigma])$, and the two types can be shown equal by $[\circ]$ and $\triangleright\beta_1$, so in intensional type theory we would need to transport one side.

Writing out the whole signature with explicit transports is difficult. The number of transports rapidly increases as later equations need to refer to transported previous types, and we may also need to introduce more transports just to rearrange previous transports over different equations. In fact, the current authors have not succeeded at writing out the type of the J[] substitution rule in intensional style. This illustrates the issue of explicit conversion derivations, which we previously explained in Section 1.1.

## 3 The Standard Model and Shallow Embedding

Previously, we described the notion of signature for the object theory, but as we remarked, merely writing down the signature in Agda is already impractical. Fortunately, we do not necessarily need the full intensional signature to be able to work with models of the object theory. The reason is that some equations can hold definitionally in specific models, thereby cutting down on the amount of transporting required. For example, if [∘] and ▹β$_1$ hold definitionally in a model, then the type of ▹β$_2$ need not include any transports.

The *standard model* of the object theory in Agda has the property that *all* of its equations hold definitionally. It was described previously by Altenkirch and Kaposi [3] similarly to the current presentation, although for a much smaller object theory.

Before presenting the model, we explain a departure from the signature described in Section 2.3. In the signature, we used natural numbers as universe levels, but in Agda, it is more convenient to use universe polymorphism and native universe levels instead. Hence, the types of the Con, Ty, Tm and Sub components become as follows:

```
Con : (i : Level) → Set (suc i)
Ty  : (j : Level) → Con i → Set (i ⊔ suc j)
Sub : Con i → Con j → Set (i ⊔ j)
Tm  : (Γ : Con i) → Ty j Γ → Set (i ⊔ j)
```

Instead of using level polymorphism, we could have used the types given in Figure 1a together with an ℕ-indexed inductive-recursive universe hierarchy, which can be implemented inside $\mathsf{Set}_0$ in Agda [19]. This choice would have added some boilerplate to the model. We choose now the more convenient version, but we note that the metatheory of universe polymorphism and universe polymorphic algebraic signatures should be investigated in future work.

### 3.1 The Standard Model

We present excerpts from the Agda formalisation, making some quantification implicit to improve readability. Let us first look at the interpretation of the type constructors of the object theory:

```
Con : (i : Level) → Set (suc i)
Con i = Set i

Ty : (j : Level) → Con i → Set (i ⊔ suc j)
Ty j Γ = Γ → Set j

Sub : Con i → Con j → Set (i ⊔ j)
Sub Γ Δ = Γ → Δ

Tm : (Γ : Con i) → Ty j Γ → Set (i ⊔ j)
Tm Γ A = (γ : Γ) → A γ
```

Contexts are interpreted as types, dependent types as type families, substitutions and terms as functions. Type and term substitution and substitution composition can be all implemented as (dependent) function composition.

```
_∘_ : Sub Θ Δ → Sub Γ Θ → Sub Γ Δ
σ ∘ δ = λ γ → σ (δ γ)

_[_] : Ty j Δ → Sub Γ Δ → Ty j Γ
A [ σ ] = λ γ → A (σ γ)

_[_] : Tm Δ A → (σ : Sub Γ Δ) → Tm Γ (A [ σ ])
t [ σ ] = λ γ → t (σ γ)
```

The empty context becomes the unit type, context extension and substitution extension are interpreted using the meta-level Σ-type.

```
• : Con zero
• = m.⊤

ε : Sub Γ •
ε = λ γ → m.tt

_▷_   : (Γ : Con i) → Ty j Γ → Con (i ⊔ j)
Γ ▷ A = m.Σ Γ A

_,_ : (σ : Sub Γ Δ) → Tm Γ (A [ σ ]) → Sub Γ (Δ ▷ A)
σ , t = λ γ → (σ γ m., t γ)

p : Sub (Γ ▷ A) Γ
p = m.fst

q : Tm (Γ ▷ A) (A [ p ])
q = m.snd
```

We interpret object-level universes with meta-level universes at the same level. Since Agda implements Russell-style universes, coding and decoding are trivial, and `Tm Γ (U j) ≡ Ty j Γ` holds definitionally in the model.

```
U : (j : Level) → Ty (suc j) Γ
U j = λ γ → Set j

El : Tm Γ (U j) → Ty j Γ
El a = a

c : Ty j Γ → Tm Γ (U j)
c A = A
```

For Π, Σ, Bool and Id, the interpretation likewise maps object-level constructions directly to their meta-level counterparts; see the formalisation [30] for details. We note here only the J[] component: its type and definition are trivial here thanks to the lack of transports. Below, `σ ↑ A` refers to the lifting of `σ : Sub Θ Γ` to `Sub (Θ ▷ A [ σ ]) (Γ ▷ A)`.

```
J[] : J C w t [ σ ]
    ≡ J (C [σ ↑ A ↑ Id (A [ p ]) (u [ p ]) q ]) (w [ σ ]) (t [ σ ])
J[] = m.refl
```

### 3.2 Shallow Embedding

Having access in Agda to the standard model of the object theory, we may now form expressions built out of model components, for example, we may define a polymorphic identity function as follows. Here, `v⁰` and `v¹` are shorthands for de Bruijn indices.

```
idfun : Tm • (Π (U zero) (Π (El v⁰) (El v¹)))
idfun = lam (lam v⁰)
```

The basic idea of shallow embedding is to view expressions such as `idfun` and its type, which are built from components of the standard model, as standing for expressions coming from an arbitrary model. This arbitrary model is often meant to be the syntax, but it does not necessarily have to be.

With `idfun`, we can enjoy the benefits of reflected equalities: we can write down `Π (El v⁰) (El v¹)` without transports, because the types of `vⁿ` de Bruijn indices compute by definition to `U zero` from `U zero [ pⁿ ]`.

A larger example for shallow embedding is presented in Section 7.2: there we prove canonicity by induction on the syntax, but represent the syntax shallowly, so we never have to prove anything about syntactic definitional equalities. Other examples are *syntactic models* [8]: this means that we build a model of an object theory from the syntax of another object theory. Every such model yields, by initiality of the syntax, a syntactic translation. We also present in Section 7.1 a formalisation of a syntactic parametricity translation in this style, using the same shallowly embedded theory for both the source and target syntaxes.

However, "pretending" that embedded expressions come from arbitrary models is only valid if we:

1. Do not construct more contexts, substitutions, terms or types than what are constructible in the syntax.
2. Do not prove more equations than what are provable about the syntax.

We will expand on the first concern in Section 6. With regards to the second concern, it would be addressed comprehensively with a proof that the standard model is *injective*. We define its statement as follows. Assume that we have a deeply embedded syntax for the object theory in Agda, with components named as `Con`, `Sub` and so on. By initiality of the syntax, there is a model morphism from the syntax to the standard model, which includes as components the following interpretation functions:

```
⟦_⟧ : Con i → Set i
⟦_⟧ : Ty j Γ → ⟦ Γ ⟧ → Set j
⟦_⟧ : Sub Γ Δ → ⟦ Γ ⟧ → ⟦ Δ ⟧
⟦_⟧ : Tm Γ A → (γ : ⟦ Γ ⟧) → ⟦ A ⟧ γ
```

Injectivity may refer to these functions; for example, injectivity on terms is stated as follows:

```
⟦⟧-injective : (t u : Tm Γ A) → ⟦ t ⟧ ≡ ⟦ u ⟧ → t ≡ u
```

However, we can show by reasoning external to Agda that injectivity of the standard model is not provable.

**Theorem 1.** *The injectivity of the standard model is not provable in Agda.*

*Proof.* We note that the object syntax includes functions which are definitionally inequal but equal extensionally, such as the following two functions:

```
f : Tm • (Π Bool Bool)
f = lam (if Bool true false v⁰)

g : Tm • (Π Bool Bool)
g = lam v⁰
```

If function extensionality is available in the metatheory, the `⟦ f ⟧` and `⟦ g ⟧` interpretations of these terms can be proven to be propositionally equal. Therefore, injectivity of the standard model and function extensionality are incompatible. But since we know that MLTT is consistent with function extensionality, it follows that injectivity of the standard model is not provable. □

This shows that the internal statement of injectivity is too strong. We weaken it by considering injectivity up to Agda's definitional equality. This requires us to step outside Agda and reason about its syntax.

### 3.3 An External View of the Standard Model

Let us consider some computation rules for the interpretation function of the standard model:

```
⟦ • ⟧         = m.⊤
⟦ Γ ▷ A ⟧     = m.Σ ⟦ Γ ⟧ ⟦ A ⟧
⟦ id ⟧        = λ γ → γ
⟦ σ ∘ δ ⟧     = λ γ → ⟦ σ ⟧ (⟦ δ ⟧ γ)
⟦ ε ⟧         = λ γ → m.tt
⟦ σ , t ⟧     = λ γ → (⟦ σ ⟧ γ m., ⟦ t ⟧ γ)
⟦ A [ σ ] ⟧ = λ γ → ⟦ A ⟧ (⟦ σ ⟧ γ)
⟦ t [ σ ] ⟧ = λ γ → ⟦ t ⟧ (⟦ σ ⟧ γ)
⟦ U j ⟧       = λ γ → Set j
...
```

If we consider the results of the interpretation function from the "outside", we see that interpreted object-theoretic terms evaluate to closed Agda terms. For example, if we have a context in the object theory:

```
Γ = • ▷ Bool ▷ Bool
```

Its `⟦ Γ ⟧` interpretation evaluates to the following closed Agda term (a left-nested Σ-type):

```
m.Σ (m.Σ m.⊤ (λ γ → m.Bool)) (λ γ → m.Bool)
```

Hence, externally, the interpretation function implements a syntactic translation which converts any object-theoretic construction to a closed Agda term. We model shallow embedding as this syntactic translation: whenever we write a shallowly embedded expression like `lam (if Bool true false v⁰)`, there is a corresponding expression in the object theory with the same shape, but in Agda this expression can be evaluated further by unfolding the definitions of the standard model.

In the next section we formalise this syntactic translation, and in Section 5 we additionally prove that it is injective. From this it follows that shallow embedding does not introduce new definitional equalities.

## 4 The Termification of a Model

For any given model $\mathcal{M} = (\mathsf{Con}, \mathsf{Ty}, \mathsf{Sub}, \mathsf{Tm}, \ldots)$ of the object type theory, we can construct a new model $\mathcal{T}^{\mathcal{M}} = (\mathsf{Con}_{\mathcal{T}}, \mathsf{Ty}_{\mathcal{T}}, \mathsf{Sub}_{\mathcal{T}}, \mathsf{Tm}_{\mathcal{T}}, \ldots)$. We call $\mathcal{T}^{\mathcal{M}}$ the *termification* of $\mathcal{M}$. The idea is that every context, type, substitution, and term can be regarded as a very specific term in the empty context; and all operations can be seen as operations on these terms.

If we take $\mathcal{M}$ to be the syntax, by initiality we get a morphism to $\mathcal{T}^{\mathcal{M}}$, which we use to model shallow embedding as a syntactic translation. Note that this

$$
\begin{aligned}
\mathsf{Con}_{\mathcal{T}}\, i &:= \mathsf{Tm}\,\bullet\,(\mathsf{U}\, i)\\
\mathsf{Ty}_{\mathcal{T}}\, j\, \Gamma &:= \mathsf{Tm}\,\bullet\,(\underline{\Gamma} \Rightarrow (\mathsf{U}\, j))\\
\mathsf{Sub}_{\mathcal{T}}\, \Gamma\, \Delta &:= \mathsf{Tm}\,\bullet\,(\underline{\Gamma} \Rightarrow \underline{\Delta})\\
\mathsf{Tm}_{\mathcal{T}}\, \Gamma\, A &:= \mathsf{Tm}\,\bullet\,(\Pi\, \underline{\Gamma}\, \underline{\mathsf{app}\, A})\\
\mathsf{id}_{\mathcal{T}} &:= \mathsf{lam}\, \mathsf{v}^0\\
\sigma \circ_{\mathcal{T}} \delta &:= \mathsf{lam}\, (\sigma[\epsilon]\, \$ (\delta[\epsilon]\, \$\, \mathsf{v}^0))\\
A[\sigma]_{\mathcal{T}} &:= \mathsf{lam}\, ((\mathsf{app}\, A)[\epsilon, \mathsf{app}\, (\sigma[\epsilon])])\\
t[\sigma]_{\mathcal{T}} &:= \mathsf{lam}\, ((\mathsf{app}\, t)[\epsilon, \mathsf{app}\, (\sigma[\epsilon])])\\
\bullet_{\mathcal{T}} &:= \mathsf{c}\, \top\\
\epsilon_{\mathcal{T}} &:= \mathsf{lam}\, \mathsf{tt}\\
\Gamma \rhd_{\mathcal{T}} A &:= \mathsf{c}\, (\Sigma\, \underline{\Gamma}\, \underline{\mathsf{app}\, A})\\
\sigma ,_{\mathcal{T}} t &:= \mathsf{lam}\, ((\mathsf{app}\, \sigma), (\mathsf{app}\, t))\\
\mathsf{p}_{\mathcal{T}} &:= \mathsf{lam}\, (\mathsf{fst}\, \mathsf{v}^0)\\
\mathsf{q}_{\mathcal{T}} &:= \mathsf{lam}\, (\mathsf{snd}\, \mathsf{v}^0)\\
\Pi_{\mathcal{T}}\, A\, B &:= \mathsf{lam}\, \big(\mathsf{c}\, (\Pi\, \underline{\mathsf{app}\, A}\, \underline{\mathsf{app}\, B[\epsilon, (\mathsf{v}^1, \mathsf{v}^0)]})\big)\\
\mathsf{lam}_{\mathcal{T}}\, t &:= \mathsf{lam}\, (\mathsf{lam}\, (t[\epsilon]\, \$ (\mathsf{v}^1, \mathsf{v}^0)))\\
\mathsf{app}_{\mathcal{T}}\, t &:= \mathsf{lam}\, (t[\epsilon]\, \$\, \mathsf{fst}\, \mathsf{v}^0\, \$\, \mathsf{snd}\, \mathsf{v}^0)\\
\Sigma_{\mathcal{T}}\, A\, B &:= \mathsf{lam}\, \big(\mathsf{c}\, (\Sigma\, \underline{\mathsf{app}\, A}\, \underline{\mathsf{app}\, B[\epsilon, (\mathsf{v}^1, \mathsf{v}^0)]})\big)\\
u ,_{\mathcal{T}} v &:= \mathsf{lam}\, (\mathsf{app}\, u, \mathsf{app}\, v)\\
\mathsf{fst}_{\mathcal{T}}\, t &:= \mathsf{lam}\, (\mathsf{fst}\, (\mathsf{app}\, t))\\
\mathsf{snd}_{\mathcal{T}}\, t &:= \mathsf{lam}\, (\mathsf{snd}\, (\mathsf{app}\, t))\\
\top_{\mathcal{T}} &:= \mathsf{lam}\, (\mathsf{c}\, \top)\\
\mathsf{tt}_{\mathcal{T}} &:= \mathsf{lam}\, \mathsf{tt}\\
\mathsf{U}_{\mathcal{T}} &:= \mathsf{lam}\, (\mathsf{c}\, (\mathsf{U}\, i))\\
\underline{a}_{\mathcal{T}} &:= a\\
\mathsf{c}_{\mathcal{T}}\, A &:= A\\
\mathsf{Bool}_{\mathcal{T}} &:= \mathsf{lam}\, (\mathsf{c}\, \mathsf{Bool})\\
\mathsf{true}_{\mathcal{T}} &:= \mathsf{lam}\, \mathsf{true}\\
\mathsf{false}_{\mathcal{T}} &:= \mathsf{lam}\, \mathsf{false}\\
\mathsf{if}_{\mathcal{T}}\, C\, u\, v\, t &:= \mathsf{lam}\, \big(\mathsf{if}\, \underline{C[\epsilon]\, \$ (\mathsf{v}^1, \mathsf{v}^0)}\, (\mathsf{app}\, u)\, (\mathsf{app}\, v)\, (\mathsf{app}\, t)\big)\\
\mathsf{Id}_{\mathcal{T}}\, A\, u\, v &:= \mathsf{lam}\, \big(\mathsf{c}\, \big(\mathsf{Id}\, \underline{\mathsf{app}\, A}\, (\mathsf{app}\, u)\, (\mathsf{app}\, v)\big)\big)\\
\mathsf{refl}_{\mathcal{T}}\, u &:= \mathsf{lam}\, (\mathsf{refl}\, (\mathsf{app}\, u))\\
\mathsf{J}_{\mathcal{T}}\, C\, w\, e &:= \mathsf{lam}\, \big(\mathsf{J}\, \underline{C[\epsilon]\, \$ (\mathsf{v}^2, \mathsf{v}^1, \mathsf{v}^0)}\, (\mathsf{app}\, w)\, (\mathsf{app}\, e)\big)
\end{aligned}
$$

**Fig. 2.** The termification construction

translation formally goes *from the object theory to the object theory.* This means that we reuse the object theory to formalise the relevant syntactic fragment of Agda. This is a fairly strong simplifying assumption, which relies on Agda conforming to the CwF formulation of type theory. However, it is also necessary, because formalising the actual implementation of Agda is not feasible.

Although our main interest is the termification of the syntax, the construction works for arbitrary models, so we present it in this generality.

The four sorts of the new model $\mathcal{T}^{\mathcal{M}}$ are the following:

$$\begin{aligned}
\mathsf{Con}_{\mathcal{T}}\,i &:= \mathsf{Tm}\,\bullet\,(\mathsf{U}\,i)\\
\mathsf{Ty}_{\mathcal{T}}\,j\,\Gamma &:= \mathsf{Tm}\,\bullet\,(\underline{\Gamma}\ \Rightarrow\ (\mathsf{U}\,j))\\
\mathsf{Sub}_{\mathcal{T}}\,\Gamma\,\Delta &:= \mathsf{Tm}\,\bullet\,(\underline{\Gamma}\ \Rightarrow\ \underline{\Delta})\\
\mathsf{Tm}_{\mathcal{T}}\,\Gamma\,A &:= \mathsf{Tm}\,\bullet\,(\Pi\,\underline{\Gamma}\,\underline{\mathsf{app}\,A})
\end{aligned}$$

All contexts, types, substitutions, and terms of the new model $\mathcal{T}^{\mathcal{M}}$ are $\mathcal{M}$-terms in the empty $\mathcal{M}$-context. It is not hard to see that the definitions above type-check: for example, if we have $\Gamma : \mathsf{Con}_{\mathcal{T}}\,i$ and $A : \mathsf{Ty}_{\mathcal{T}}\,j\,\Gamma$, then by definition $\underline{\Gamma} : \mathsf{Ty}\,i\,\bullet$ and $\underline{\mathsf{app}\,A} : \mathsf{Ty}\,j\,(\bullet \rhd \underline{\Gamma})$, which means we can build $\Pi\,\underline{\Gamma}\,\underline{\mathsf{app}\,A}$ as in the definition of $\mathsf{Tm}_{\mathcal{T}}\,\Gamma\,A$.

The object theory, as shown in Figures 1a and 1b, has 29 operators. In Figure 2, we show how all 29 operators (together with the four sorts) of the model $\mathcal{T}^{\mathcal{M}}$ are constructed from components of $\mathcal{M}$. Finally, it is straightforward albeit tedious to check the 37 equalities that are required to hold. We have done the calculations both with pen and paper and in Agda. We do not give explicit paper proofs, but we refer to our formalisation instead: there, we state all equalities explicitly, and they are all proved using `m.refl`. This concludes the construction of the model $\mathcal{T}^{\mathcal{M}}$.

## 5 The Injectivity Result

In this section, we show that we can shallowly embed the syntax without creating new definitional equalities.

If we apply the termification construction of Section 4 on the syntax $\mathsf{Syn}$, we get a model $\mathcal{T}^{\mathsf{Syn}}$. Further, we have a morphism of models $[\![\cdot]\!] : \mathsf{Syn} \to \mathcal{T}^{\mathsf{Syn}}$ by the initiality of the syntax which maps $\bullet\ :\ \mathsf{Con}\,0$ to $[\![\bullet]\!] = \bullet_{\mathcal{T}}$, and which maps $\Gamma \rhd A : \mathsf{Con}\,i$ to $[\![\Gamma \rhd A]\!] = [\![\Gamma]\!] \rhd_{\mathcal{T}} [\![A]\!]$, and so on.

An interesting property of the morphism $[\![\cdot]\!]$ is that it is *injective.* Before stating precisely what this means, we need the following definition:

**Definition 1.** *Given two contexts* $\Gamma : \mathsf{Con}\,i$*,* $\Delta : \mathsf{Con}\,j$ *in the object theory [or any model* $\mathcal{M}$*], we write* $\Gamma \simeq \Delta$ *for the type in the metatheory whose elements are quadruples* $F = (F_1, F_2, F_{12}, F_{21})$ *as follows:* $F_1$ *and* $F_2$ *are substitutions in the syntax [more generally, in* $\mathcal{M}$*] and* $F_{12}$*,* $F_{21}$ *are equalities,*

$$F_1\ : \mathsf{Sub}\,\Gamma\,\Delta$$

$$\begin{aligned} F_2 \;&: \mathsf{Sub}\,\Delta\,\Gamma \\ F_{12} &: F_2 \circ F_1 = \mathsf{id}_\Gamma \\ F_{21} &: F_1 \circ F_2 = \mathsf{id}_\Delta. \end{aligned}$$

*We call such a quadruple an* isomorphism.

**Theorem 2.** *The morphism of models* $[\![\cdot]\!] : \mathsf{Syn} \to \mathcal{T}^{\mathsf{Syn}}$ *is injective, in the following sense:*

*(T1) If* $\Gamma : \mathsf{Con}\,i$, $\Delta : \mathsf{Con}\,j$ *are contexts such that* $[\![\Gamma]\!] = [\![\Delta]\!]$*, then we have* $\Gamma \simeq \Delta$.
*(T2) If* $A, B : \mathsf{Ty}\,i\,\Gamma$ *are types such that* $[\![A]\!] = [\![B]\!]$*, then we have* $A = B$.
*(T3) If* $\sigma, \tau : \mathsf{Sub}\,\Gamma\,\Delta$ *are substitutions such that* $[\![\sigma]\!] = [\![\tau]\!]$*, then* $\sigma = \tau$.
*(T4) If* $s, t : \mathsf{Tm}\,\Gamma\,A$ *are terms such that* $[\![s]\!] = [\![t]\!]$*, then we have* $s = t$.

*Proof.* We show the following metatheoretic statements:

(P1) For a context $\Gamma : \mathsf{Con}\,i$, we have an element $(\Gamma_1, \Gamma_2, \Gamma_{12}, \Gamma_{21})$ of

$$\Gamma \simeq \left(\bullet \triangleright \underline{\underline{[\![\Gamma]\!]}}\right)$$

(P2) For a type $A : \mathsf{Ty}\,i\,\Gamma$, we have an equation

$$A_= :\ A \;=\; \underline{\mathsf{app}\,[\![A]\!]}[\Gamma_1]$$

(P3) For a substitution $\sigma : \mathsf{Sub}\,\Gamma\,\Delta$, we have an equation

$$\sigma_= :\ \sigma \;=\; \Delta_2 \circ (\epsilon, \mathsf{app}\,[\![\sigma]\!]) \circ \Gamma_1$$

(P4) For a term $t : \mathsf{Tm}\,\Gamma\,A$, we have an equation

$$t_= :\ t \;=\; (\mathsf{app}\,[\![t]\!])[\Gamma_1]$$

Of course, the statement of the theorem follows easily from (P1)–(P4); for example, if we have $[\![s]\!] = [\![t]\!]$ as in (T4), we get $s = (\mathsf{app}\,[\![s]\!])[\Gamma_1] = (\mathsf{app}\,[\![t]\!])[\Gamma_1] = t$ from the above.

Before verifying (P1)–(P4), we can first convince ourselves that these expressions type-check in the extensional type theory which we use as metatheory. For (P1), this is clear. In (P2), the types are as follows:

$$\begin{array}{lll} & A & : \mathsf{Ty}\,i\,\Gamma \\ \textit{thus} & [\![A]\!] & : \mathsf{Tm}\,\bullet\,(\underline{\underline{[\![\Gamma]\!]}} \Rightarrow \mathsf{U}\,i) \\ \textit{thus} & \mathsf{app}\,[\![A]\!] & : \mathsf{Tm}\,(\bullet \triangleright \underline{\underline{[\![\Gamma]\!]}})\,\mathsf{U}\,i \\ \textit{thus} & \underline{\mathsf{app}\,[\![A]\!]} & : \mathsf{Ty}\,i\,(\bullet \triangleright \underline{\underline{[\![\Gamma]\!]}}) \\ \textit{thus} & \underline{\mathsf{app}\,[\![A]\!]}[\Gamma_1] & : \mathsf{Ty}\,i\,\Gamma \end{array}$$

The case (P4) is almost identical to this, but needs to make use of (P2):

$$\begin{array}{ll} t & : \mathsf{Tm}\,\Gamma\,A \end{array}$$

$$\begin{array}{lll} \textit{thus} & [\![t]\!] & : \mathsf{Tm}\,\bullet\,(\Pi\,\underline{[\![\Gamma]\!]}\,\underline{\mathsf{app}\,[\![A]\!]}) \\ \textit{thus} & \mathsf{app}\,[\![t]\!] & : \mathsf{Tm}\,(\bullet \triangleright \underline{[\![\Gamma]\!]})\,\underline{\mathsf{app}\,[\![A]\!]} \\ \textit{thus} & (\mathsf{app}\,[\![t]\!])[\Gamma_1] & : \mathsf{Tm}\,\Gamma\,(\underline{\mathsf{app}\,[\![A]\!]}[\Gamma_1]) \\ \textit{by } A_= & (\mathsf{app}\,[\![t]\!])[\Gamma_1] & : \mathsf{Tm}\,\Gamma\,A \end{array}$$

One checks similarly that (P3) type-checks.

We prove (P1)–(P4) by constructing a displayed model. As described in Section 2.3, this corresponds to "induction over the syntax".

To construct the displayed model, we need to cover the four sorts, 29 operators, and 37 equalities in Figures 1a and 1b. The components for the four sorts are given by (P1)–(P4). Two of the 29 operators construct a context, namely $\bullet$ and $\triangleright$; for these, we need to construct an isomorphism. For the remaining 27 operators, we need to prove an equality. The components for the 37 equalities are automatic: Since (P2)–(P4) are equalities, all equality components of the displayed model amount to equalities between equalities, which are trivial in our extensional metatheory. Note that none of the equalities in Figures 1a and 1b are between contexts.

We start with the two operators that construct contexts. The case for the empty context is easy: we need to find $(\bullet_1, \bullet_2, \bullet_{12}, \bullet_{21})$ showing

$$\bullet \;\simeq\; (\bullet \triangleright \underline{[\![\bullet]\!]})$$

This is simple:

$$\begin{array}{rl} \bullet_1 & : \;\; \mathsf{Sub}\,\bullet\,(\bullet \triangleright \underline{[\![\bullet]\!]}) \\ \bullet_1 & := (\epsilon, \mathsf{tt}) \\ \bullet_2 & : \;\; \mathsf{Sub}\,(\bullet \triangleright \underline{[\![\bullet]\!]})\,\bullet \\ \bullet_2 & := \epsilon \end{array}$$

The equality $\bullet_{12}$ follows from $\bullet\eta$, and the equality $\bullet_{21}$ follows from $\triangleright\eta$ and $\top\eta$.

Next, we have the case $\Gamma \triangleright A$, where we can already assume the property (P1) for $\Gamma$ and (P2) for $A$. After unfolding the definition of $[\![\Gamma \triangleright A]\!] = [\![\Gamma]\!] \triangleright_{\mathcal{T}} [\![A]\!]$, we see that we have to construct an isomorphism

$$(\Gamma \triangleright A) \;\simeq\; \big(\bullet \triangleright \Sigma\,\underline{[\![\Gamma]\!]}\,\underline{\mathsf{app}\,[\![A]\!]}\big)$$

The two substitutions are:

$$\begin{array}{rl} (\Gamma \triangleright A)_1 & : \;\; \mathsf{Sub}\,(\Gamma \triangleright A)\,\big(\bullet \triangleright \Sigma\,\underline{[\![\Gamma]\!]}\,\underline{\mathsf{app}\,[\![A]\!]}\big) \\ (\Gamma \triangleright A)_1 & := \big(\epsilon, (\mathsf{v}^0[\Gamma_1 \circ \mathsf{p}], \mathsf{v}^0)\big) \\ (\Gamma \triangleright A)_2 & : \;\; \mathsf{Sub}\,\big(\bullet \triangleright \Sigma\,\underline{[\![\Gamma]\!]}\,\underline{\mathsf{app}\,[\![A]\!]}\big)\,(\Gamma \triangleright A) \\ (\Gamma \triangleright A)_2 & := \big(\Gamma_2 \circ (\epsilon, \mathsf{fst}\,\mathsf{v}^0), \mathsf{snd}\,\mathsf{v}^0\big) \end{array}$$

Quick calculations give us

$$(\Gamma \triangleright A)_1 \circ (\Gamma \triangleright A)_2$$

$$
\begin{aligned}
&= \big(\epsilon, (\mathsf{v}^0[\Gamma_1 \circ \mathsf{p}], \mathsf{v}^0)\big) \circ \big(\Gamma_2 \circ (\epsilon, \mathsf{fst}\,\mathsf{v}^0), \mathsf{snd}\,\mathsf{v}^0\big) \\
&= \big(\epsilon, (\mathsf{v}^0[\Gamma_1 \circ \Gamma_2 \circ (\epsilon, \mathsf{fst}\,\mathsf{v}^0)], \mathsf{snd}\,\mathsf{v}^0)\big) \\
&= \big(\epsilon, (\mathsf{fst}\,\mathsf{v}^0, \mathsf{snd}\,\mathsf{v}^0)\big) \\
&= \big(\epsilon, \mathsf{v}^0\big) \\
&= (\mathsf{p}, \mathsf{q}) \\
&= \mathsf{id}
\end{aligned}
$$

as well as

$$
\begin{aligned}
&(\Gamma \triangleright A)_2 \circ (\Gamma \triangleright A)_1 \\
&= \big(\Gamma_2 \circ (\epsilon, \mathsf{fst}\,\mathsf{v}^0), \mathsf{snd}\,\mathsf{v}^0\big) \circ \big(\epsilon, (\mathsf{v}^0[\Gamma_1 \circ \mathsf{p}], \mathsf{v}^0)\big) \\
&= \big(\Gamma_2 \circ (\epsilon, \mathsf{v}^0[\Gamma_1 \circ \mathsf{p}]), \mathsf{v}^0\big) \\
&= \big(\Gamma_2 \circ ((\mathsf{p}, \mathsf{q}) \circ (\Gamma_1 \circ \mathsf{p})), \mathsf{v}^0\big) \\
&= \big(\Gamma_2 \circ \Gamma_1 \circ \mathsf{p}, \mathsf{v}^0\big) \\
&= (\mathsf{p}, \mathsf{q}) \\
&= \mathsf{id}
\end{aligned}
$$

The first of the remaining 27 operations is the identity substitution $\mathsf{id} : \mathsf{Sub}\,\Gamma\,\Gamma$, where we can already assume property (P1) for $\Gamma$. We need to show

$$\mathsf{id}_= : \mathsf{id} = \Gamma_2 \circ (\epsilon, \mathsf{app}\,[\![\mathsf{id}]\!]) \circ \Gamma_1$$

We unfold $[\![\mathsf{id}]\!] = \mathsf{id}_{\mathcal{T}} = \mathsf{lam}\,\mathsf{v}^0$ and use $\Pi\eta$ to simplify the right-hand side of the equation to

$$\Gamma_2 \circ (\epsilon, \mathsf{v}^0) \circ \Gamma_1,$$

which by $\bullet\eta$, $\triangleright\eta$ and $\Gamma_{12}$ is equal to $\mathsf{id}$ as required.

The calculations for the remaining 26 operations are similar, Appendix A contains all of them in full detail. For completeness, the components discussed above are included in the figure as well. This completes the proof of the injectivity result. □

## 6 Wrapped Standard Model

In the previous section, we have shown that our specific version of shallow embedding does not introduce new definitional equalities. However, in practice we can only apply Theorem 2 if there actually exists an object-theoretic expression which is embedded, but there are many inhabitants in the standard model which do not arise as interpretations of object-theoretic expressions.

For example, contexts are interpreted as left-nested Σ-types, but since `Con i` is defined as `Set i` in the standard model, we can just inhabit `Con zero` with `m.Bool` or any small Agda type. This would be morally incorrect in a shallow embedding situation, since we might rely on properties that are not provable about the object syntax.

Additionally, even if we avoid extraneous inhabitants, some propositional equalities may be provable in the standard model, which are provable false in the syntax. In Proof 1 we gave such an example, where function extensionality yields additional equality proofs. In general, we want the freedom to assume function extensionality and other extensionality principles (e.g. for propositions or coinductive types) in the metatheory, so outlawing these principles in the metatheory is not acceptable as an enforcer of moral conduct.

Our proposed enforcement method is the following: wrap the interpretations of contexts, terms, substitutions and types in the standard model in unary record types, whose constructors are private and thus invisible to external modules. For contexts and types, the wrappers are as follows:

```
record Con' i : Set (suc i) where
  constructor mkC
  field
    |_|C : Set i

record Ty' (j : Level)(Γ : Con' i) : Set (i ⊔ suc j) where
  constructor mkT
  field
    |_|T : | Γ |C → Set j
```

We define `Sub'` and `Tm'` likewise, with `mks`, `|_|s`, `mkt` and `|_|t`, and put these four types in a module. In a different module, we define the "wrapped" standard model. The sorts in the model are defined using the wrapper types:

```
Con : (i : Level) → Set (suc i)
Con = Con'

Ty : (j : Level)(Γ : Con i) → Set (i ⊔ suc j)
Ty = Ty'

Sub : Con i → Con j → Set (i ⊔ j)
Sub = Sub'

Tm : (Γ : Con i) → Ty j Γ → Set (i ⊔ j)
Tm = Tm'
```

The rest of the model needs to be annotated with wrapping and unwrapping. Some examples for definitions, omitting type declarations for brevity:

```
id      = mks λ γ → γ
σ ∘ δ   = mks λ γ → | σ |s (| δ |s γ)
A [ σ ] = mkT λ γ → | A |T (| σ |s γ)
t [ σ ] = mkt λ γ → | t |t (| σ |s γ)
•       = mkC m.⊤
ε       = mks λ γ → m.tt
```

```
Γ ▷ A    = mkC (m.Σ | Γ |C | A |)
σ , t    = mks λ γ → (| σ |s γ m.,Σ | t |t γ)
p        = mks m.fst
q        = mkt m.snd
U j      = mkT λ γ → Set j
El a     = mkT | a |t
c A      = mkt | A |T
```

Importantly, the wrapped model still *supports all equations definitionally*. This is possible because the wrapper record types support $\eta$-equality, which expresses that `mkC | Γ |C` is definitionally equal to `Γ`, and likewise for the other wrappers. In short, unary records in Agda yield isomorphisms of types up to definitional equality.

The usage of the wrapped standard model for shallow embedding is simply as follows: we import the wrapped standard model, but do not import the module containing the wrapper types.

This way, there is no way to refer to the internals of the model. In fact, the only way to construct any inhabitants of the embedded syntax in this setup is to explicitly refer to the components of the wrapped model. For instance, `Con zero` cannot be anymore inhabited with `m.Bool`, since `m.Bool` has type `Set₀`, but we need a `Con' zero`, which we can only inhabit now using the empty context and context extension.

## 7 Case Studies

As a demonstration of using the shallowly embedded syntax, in this section we describe our formalisation of a syntactic parametricity translation and a canonicity proof for MLTT. These are formalised as displayed models over the syntax (that is, over the wrapped standard model described in Section 6).

### 7.1 Parametricity

Parametricity was introduced by Reynolds [41] in order to formalise the notion of representation independence. The unary version of his parametricity theorem states that terms preserve logical predicates: if a predicate holds for a semantic context, then it holds for the interpretation of the term at that context. Reynolds formulated parametricity as a model construction of System F. Bernardy et al. [6] noticed that type theory is powerful enough to express statements about its own parametricity and defined parametricity as a syntactic operation. This operation turns a context into a lifted context which has a witness of the logical predicate for each type in the original context. There is a projection from this lifted context back to the original context. A type $A$ is turned into a predicate over $A$ in the lifted context and a term is turned into a witness of the predicate for its type in the lifted context. We note that a more indexed version of this translation can be defined: This turns a context is into a type in the original context (that is, a predicate over the original context), a type into a predicate over the original context, a witness of the predicate for the original context and an element of the type. Substitutions and terms are turned into terms expressing preservation of the predicates. We define this indexed version of the translation in Agda.

The sorts are given as follows in our displayed model. We use `S.` prefixes to refer to the syntax, and use `-ˢ` superscripts on variables coming from the syntax.

```
Con : ∀ i → S.Con i → Set (suc i)
Con i Γˢ = S.Ty i Γˢ

Ty : ∀ i (Γ : Con j Γˢ) (Aˢ : S.Ty i Γˢ) → Set (suc i ⊔ j)
Ty i Γ Aˢ = S.Ty i (Γˢ S.▷ Γ S.▷ Aˢ S.[ S.p ])

Sub : ∀ (Γ : Con i Γˢ)(Δ : Con j Δˢ) → S.Sub Γˢ Δˢ → Set (i ⊔ j)
Sub Γ Δ σˢ = S.Tm (Γˢ S.▷ Γ) (Δ S.[ σˢ S.∘ S.p ])

Tm : ∀ (Γ : Con i Γˢ)(A : Ty j Γ Aˢ) → S.Tm Γˢ Aˢ → Set (i ⊔ j)
Tm Γ A tˢ = S.Tm (Γˢ S.▷ Γ) (A S.[ S.id S., tˢ S.[ S.p ] ])
```

A context over a syntactic context `Γˢ` is a syntactic type in `Γˢ`. A type over a syntactic type `Aˢ` is a syntactic type in the context `Γˢ` extended with two more components: `Γ`, that is the logical predicate for `Γˢ` and `Aˢ` itself (which has to be weakened using `S.p`). A substitution over `σˢ` is a term in context `Γˢ S.▷ Γ` which has a type saying that the predicate `Δ` holds for `σˢ`. We have the analogous statement for terms. We refer to the formalisation [30] for the rest of the displayed model, it follows the original parametricity translation.

All equalities of the displayed model hold definitionally. Compared to a previous formalisation using a deep embedding [3], it is significantly shorter (322 vs. 1682 lines of code – we only counted the lines of code for the substitution calculus, Π and the universe because only these were treated in the previous formalisation). Note that although we implemented the displayed model, we did not implement the corresponding eliminator function which translates an `S`-term into its interpretation; we discuss such eliminators in Section 8.2.

### 7.2 Canonicity

Canoncity for type theory states that a term of type `Bool` in the empty context is equal to either `true` or `false`. Following [14,27] this can be proven by another logical predicate argument. We formalise this logical predicate as the following displayed model. We list the definitions for sorts and `Bool` for illustration.

```
Con : ∀ i → S.Con i → Set (suc i)
Con i Γˢ = S.Sub S.∙ Γˢ → Set i

Ty  : ∀ i (Γ : Con j Γˢ) (Aˢ : S.Ty i Γˢ) → Set (suc i ⊔ j)
Ty i Γ Aˢ = ∀ {ρˢ} → Γ ρˢ → S.Tm S.∙ (Aˢ S.[ ρˢ ]) → Set i
```

```
Sub : ∀ (Γ : Con i Γˢ)(Δ : Con j Δˢ) → S.Sub Γˢ Δˢ → Set (i ⊔ j)
Sub Γ Δ σˢ = ∀ {ρˢ} → Γ ρˢ → Δ (σˢ S.∘ ρˢ)

Tm : ∀ (Γ : Con i Γˢ)(A : Ty j Γ Aˢ) → S.Tm Γˢ Aˢ → Set (i ⊔ j)
Tm Γ A tˢ = ∀ {ρˢ}(ρ' : Γ ρˢ) → A ρ' (tˢ S.[ ρˢ ])

Bool : Ty zero Γ S.Bool
Bool ρ' tˢ = m.Σ m.Bool λ β → m.if _ S.true S.false β ≡ tˢ
```

A context over $\Gamma^s$ is a proof-relevant predicate over closed substitutions into $\Gamma^s$. A type over $A^s$ is a proof-relevant predicate over closed terms of type $A$ where the type is substituted by a closed substitution for which the predicate holds. A substitution over $\sigma^s$ is a function which says that if the predicate `Γ` holds for a closed substitution $\rho^s$ then `Δ` holds for $\sigma^s$ composed with $\rho^s$. A term over $t^s$ similarly states that if `Γ` holds for a $\rho^s$, then `A` holds for `tˢ S.[ ρˢ ]`.

The predicate `Bool` holds for a closed term $t^s$ of type `S.Bool` if there is a metatheoretic boolean (`β : m.Bool`) which when converted to a syntactic boolean is equal to $t^s$: in short, it holds if $t^s$ is either `S.true` or `S.false`. The equality is expressed as a metatheoretic equality `≡`, which we generally use for representing conversion for the object syntax.

The formalisation of canonicity consists of roughly 1000 lines of Agda code. However, out of this, 400 lines are automatically generated type signatures, which are of no mathematical interest, and are necessary only because of technical problems in Agda's inference of implicit parameters. These problems also prevented us from formalising the `J[]` component in the displayed model, but otherwise the formalisation is complete.

### 7.3 Termification and Injectivity

We also implemented termification (Section 4) in Agda as a model and it is also possible to implement the injectivity proof (Section 5) using the shallow embedding, without postulating an elimination principle of the shallow syntax (the Agda proof of injectivity is not yet completed). Injectivity is given by a displayed model over the syntax which contains both the termification model of the syntax and the (P1)–(P4) components of the injectivity proof as follows. We use `TS.` prefix to refer to components of the termified model for the syntax.

```
record Con i (Γˢ : S.Con i) : Set (suc i) where
  field
    ⟦_⟧ : TS.Con i
    _₁  : S.Sub Γˢ (S.• S.▷ S.El ⟦_⟧)
    _₂  : S.Sub (S.• S.▷ S.El ⟦_⟧) Γˢ
    _₁₂ : _₁ S.∘ _₂ ≡ S.id
    _₂₁ : _₂ S.∘ _₁ ≡ S.id
```

```
record Ty j (Γ : Con i Γˢ) (Aˢ : S.Ty j Γˢ) : Set (i ⊔ suc j) where
  field
    ⟦_⟧ : TS.Ty j ⟦ Γ ⟧
    _⁼  : Aˢ ≡ S.El (S.app ⟦_⟧ S.[ Γ ₁ ])

record Sub (Γ : Con i Γˢ)(Δ : Con j Δˢ)(σˢ : S.Sub Γˢ Δˢ) :
  Set (i ⊔ j) where
  field
    ⟦_⟧ : TS.Sub ⟦ Γ ⟧ ⟦ Δ ⟧
    _⁼  : σˢ ≡ (Δ ₂ S.∘ (S.p S., S.app ⟦_⟧)) S.∘ Γ ₁

record Tm (Γ : Con i Γˢ)(A : Ty j Γ Aˢ)(tˢ : S.Tm Γˢ Aˢ) :
  Set (i ⊔ j) where
  field
    ⟦_⟧ : TS.Tm ⟦ Γ ⟧ ⟦ A ⟧
    _⁼  : tˢ ≡ m.tr (S.Tm Γˢ) (A ⁼ m.⁻¹) (S.app ⟦_⟧ S.[ Γ ₁ ])
```

The ⟦_⟧ components are just the termification model while the rest of the record types implement (P1)–(P4). Compared to the proof presented in this paper using the extensional metatheory, in Agda the last equation contains an explicit transport `m.tr` over the equality proof `A ⁼`.

# 8 Discussion

## 8.1 Range of Embeddable Object Theories

So far, we focused on a particular object theory, which was described in Section 2.3 in detail. However, there is a rather wide range of object theories suitable for shallow embedding. There are some features which the object theory must possess. We discuss these in the following in an informal way.

First, object theories must support a "standard model" in the metatheory, which is injective in the external sense described in our paper. External injectivity is important: for example, for a large class of algebraic theories, *terminal models* exist (see e.g. [29]), where every type is interpreted as the unit type. The motivation of shallow embedding is to get more definitional equalities, but in terminal models we get too much of it, because all inhabitants are definitionally equal. Injectivity filters out dubious embeddings like terminal models.

The notion of standard model is itself informal. We may say that a standard model should interpret object-level constructions with essentially the same meta-level constructions. This is clearly the case when we model type theories in Agda which are essentially syntactic fragments of Agda. However, this should not be taken rigidly, as there might be externally injective shallow embeddings which do not fall into the standard case of embedding syntactic fragments. Thus far we have not investigated such theories; this could be a potential line of future work.

Some language-like theories, although widely studied, do not seem to support shallow embedding. For example, partial programming languages do not admit a standard `Set`-interpretation; they may have other models, but those are unlikely to support useful definitional equalities, when implemented in MLTT. However, a potential future proof assistant for synthetic domain theory [7] could support useful shallow embedding for partial languages. Likewise, variants of type theories such as cubical [13] or modal type theories could present further opportunities for shallow embeddings which are not available in MLTT.

On the other hand, undecidable definitional equality in the object theory does not necessarily preclude shallow embedding. For example, we could add equality reflection to the object theory considered in this paper, thereby making its definitional equality undecidable. Assuming `funext : (∀ x → f x ≡ g x) → f ≡ g`, we can interpret equality reflection as follows in the standard model:

```
reflect : (t u : Tm Γ A) → Tm Γ (Id A t u) → t ≡ u
reflect t u p = funext p
```

So, the standard model of an extensional object theory has one equation which is not definitional anymore: the interpretation of equality reflection. But we still get all the previous benefits from the other definitional equalities in the model.

Generally, if the equational theories on the object-level and the meta-level do not match exactly, shallow embedding is still usable.

If the metatheory has **too many** definitional equalities, then we can just modify the standard model in order to eliminate the extra equalities. For example, if the object theory does not have $\eta$ for functions, we can introduce a wrapper type for functions, with $\eta$-equality turned off[1]:

```
record Π' {i}{j}(A : Set i)(B : A → Set j) : Set (i ⊔ j) where
  no-eta-equality
  constructor lam'
  field
    app' : ∀ x → B x
```

$\eta$ can be still proven for `Π'` propositionally, however using the wrapping trick (Section 6) this equality won't be exported when using the syntax.

If the metatheory has **too few** definitional equalities, then shallow embedding might still be possible with some equations holding only propositionally. We saw such an example with the shallow embedding of equality reflection. However, if we can reflect some but not all equalities, that can be still very helpful in practical formalisations.

### 8.2 Recursors and Eliminators for the Embedded Syntax

Shallow embedding gave us a particular model with strict equalities. The question is: assuming that we only did morally correct constructions, is it consistent to assume that the embedded syntax is really the syntax, i.e. it supports recursion and induction principles? For example, for our object theory, initiality (i.e.

[1] Or use an inductive type definition instead of a record.

unique recursion) for the embedded syntax means that for any other model `M` containing `Conᴹ`, `Tyᴹ`, `Subᴹ` etc. components, there is a model morphism from the embedded syntax to `M` which includes the following functions:

```
⟦_⟧ : Con i → Conᴹ i
⟦_⟧ : Ty j Γ → Tyᴹ j ⟦ Γ ⟧
...
```

If "morally correct" means that all of our constructions can be in principle translated to constructions on deeply embedded syntax, then it is clearly consistent to rely on postulated initiality. We note here that the translation from shallow to deeply embedded syntax is an instance of translating from extensional type theory to intensional type theory [21,45], which introduces transports and invocations of function extensionality in order to make up for missing definitional equalities. However, in this paper we do not investigate moral correctness more formally.

If we do postulate initiality for the embedded syntax, we should be prepared that recursors and eliminators are unlikely to compute in any current proof assistant. In Agda, we attempted to use rewrite rules to make a postulated recursor compute on shallow syntax; this could be in principle possible, but the $\beta$-rules for the recursor seem to be illegal in Agda as rewrite rules. How great limitation the lack of computing recursion is? We argue that it is not as bad as it seems.

First, in the literature for semantics of type theory, it is rare that models of type theory make essential use of recursors of other models. The only example we know is in a previous work by two of the current authors and Altenkirch [29].

Second, many apparent uses of recursors in models are not essential, and can be avoided by reformulating models. We used such a technique in Section 7.3. Here we give a much simpler analogous example: writing a sorting function for lists of numbers, in two ways:

1. First, we write a sorting function, given by the recursor for a model of the theory of lists. Then, we prove by induction on lists that the function's output is really sorted. The latter step is given by a displayed model over the syntax of lists, which displayed model refers to the previous recursor.
2. We write a function which returns a $\Sigma$-type containing a list together with a proof that it is sorted.

In the latter case, we only use a single non-displayed model, and there is no need to refer to any recursor in the model.

### 8.3 Ergonomics

We consider here the experience of using shallowing embedding in proof assistants, in particular in Agda, where we have considerable experience as users of the technique. We focus on issues and annoyances, since the benefits of shallow embedding have been previously discussed.

*Goal types and error messages* are not the best, since they all talk about expressions in the wrapped standard model instead of the deeply embedded syntax. Hence, working with shallow embedding requires us to mentally translate between syntax and the standard model. It should be possible in principle to back-translate messages to deep syntax. In Agda, `DISPLAY` pragmas can be used to display expressions in user-defined way, but it seems too limited for our purpose.

*Increased universe level of the embedded syntax.* Let us assume an object type theory without a universe hierarchy. In this case the type of contexts can be given as `Con : Set₀` in an inductive `data` definition or a postulated quotient inductive definition. In contrast, the standard model *defines* `Con` as `Set`, hence `Con` has type `Set₁` in this case. In Agda, this increase in levels can cause additional boilerplate and usage of explicit level lifting. A way to remedy this is to define `Con` as a custom inductive-recursive universe, which can usually fit into `Set₀`, but in this case we get additional clutter in system messages arising from inductive-recursive decoding.

## 9 Conclusions

In this paper, we investigated the shallow embedding of a type theory into type theory. We motivated it as an effective technique to reflect definitional equalities of an object type theory. We showed that shallow embedding of a particular object theory is really an embedding, since it is injective in an external sense.

We do not suggest that shallow embedding can replace deep embedding in every use case. For example, when implementing a type checker or compiler, one has to use deep embeddings. We hope that future proof assistants will be robust and powerful enough to allow feasible direct formalisations and make shallow embeddings unnecessary.

A potential line of future work would be to try to use shallow embedding as presented here for other object theories and formalisations. Subjectively, shallow embedding made a huge difference when we formalised our case studies; a previous formalisation [3] of the parametricity translation took the current first author months to finish, while the current formalisation took less than a day, for a much larger object theory. Formalisations which were previously too tedious to undertake could be within reach now. Also, it could be explored in the future whether morally correct shallow embedding works for object theories which are not just syntactic fragments of the metatheory. For instance, structured categories other than CwFs, such as monoidal categories could be investigated for shallow embedding.

# A The injectivity displayed model

We list the components of the displayed model for the injectivity proof described in Section 5. We don't write subscripts for metavariables and operators of the syntax, only for components of the displayed model ($_1$, $_2$, $_{12}$, $_{21}$ and $_=$).

$$\begin{aligned}
&\mathsf{Con}\, i\, \Gamma && :=\Gamma\ \simeq\ \big(\bullet \triangleright \underline{\llbracket\Gamma\rrbracket}\big)\\
&\mathsf{Ty}\, j\, (\Gamma_1,\Gamma_2,\Gamma_{12},\Gamma_{21})\, A && := A = \underline{\mathsf{app}\, \llbracket A\rrbracket}[\Gamma_1]\\
&\mathsf{Sub}\, (\Gamma_1,\Gamma_2,\Gamma_{12},\Gamma_{21})\, (\Delta_1,\Delta_2,\Delta_{12},\Delta_{21})\, \sigma && := \sigma\ =\ \Delta_2 \circ (\epsilon, \mathsf{app}\, \llbracket\sigma\rrbracket) \circ \Gamma_1
\end{aligned}$$

$$\mathsf{Tm}\,(\Gamma_1,\Gamma_2,\Gamma_{12},\Gamma_{21})\,A_=\,t \quad := t \ = \ (\mathsf{app}\,[\![t]\!])[\Gamma_1]$$

$$\begin{array}{lll}
\mathsf{id}_= & : & \mathsf{id} = \\
 & & \Gamma_2 \circ \Gamma_1 = \\
 & & \Gamma_2 \circ (\epsilon, \mathsf{app}\,(\mathsf{lam}\,\mathsf{v}^0)) \circ \Gamma_1 = \\
 & & \Gamma_2 \circ (\epsilon, \mathsf{app}\,[\![\mathsf{id}]\!]) \circ \Gamma_1 \\
\sigma_= \circ_= \delta_= & : & \sigma \circ \delta = \\
 & & \Delta_2 \circ (\epsilon, \mathsf{app}\,[\![\sigma]\!]) \circ \Theta_1 \circ \Theta_2 \circ (\epsilon, \mathsf{app}\,[\![\delta]\!]) \circ \Gamma_1 = \\
 & & \Delta_2 \circ (\epsilon, \mathsf{app}\,[\![\sigma]\!][\epsilon, \mathsf{app}\,[\![\delta]\!]]) \circ \Gamma_1 = \\
 & & \Delta_2 \circ (\epsilon, ([\![\sigma]\!][\epsilon]\,\$([\![\delta]\!][\epsilon]\,\$\,\mathsf{v}^0))) \circ \Gamma_1 = \\
 & & \Delta_2 \circ (\epsilon, \mathsf{app}\,[\![\sigma \circ \delta]\!]) \circ \Gamma_1 \\
A_=[\sigma_=]_= & : & A[\sigma] = \underline{\mathsf{app}[\![A]\!]}[\Delta_1][\Delta_2 \circ (\epsilon, \mathsf{app}\,[\![\sigma]\!]) \circ \Gamma_1] = \\
 & & \underline{(\mathsf{app}\,[\![A]\!])[\epsilon, \mathsf{app}\,([\![\sigma]\!][\epsilon])]}[\Gamma_1] = \underline{\mathsf{app}\,[\![A[\sigma]]\!]}[\Gamma_1] \\
t_=[\sigma_=]_= & : & t[\sigma] = (\mathsf{app}[\![t]\!])[\Delta_1][\Delta_2 \circ (\epsilon, \mathsf{app}\,[\![\sigma]\!]) \circ \Gamma_1] = \\
 & & (\mathsf{app}\,[\![t]\!])[\epsilon, \mathsf{app}\,([\![\sigma]\!][\epsilon])][\Gamma_1] = \mathsf{app}\,[\![t[\sigma]]\!][\Gamma_1] \\
\bullet_1 & := & (\epsilon, \mathsf{tt}) \\
\bullet_2 & := & \epsilon \\
\bullet_{12} & : & \bullet_1 \circ \bullet_2 = (\epsilon, \mathsf{tt}) \circ \epsilon = (\epsilon, \mathsf{tt}) = (\mathsf{p}, \mathsf{q}) = \mathsf{id} \\
\bullet_{21} & : & \bullet_2 \circ \bullet_1 = \epsilon \circ (\epsilon, \mathsf{tt}) = \epsilon = \mathsf{id} \\
\epsilon_= & : & \epsilon = \epsilon \circ \cdots = \bullet_2 \circ (\epsilon, \mathsf{app}\,[\![\sigma]\!]) \circ \Gamma_1 \\
(\Gamma_1, \dots) \rhd_1 A_= & := & \big(\epsilon, (\mathsf{v}^0[\Gamma_1 \circ \mathsf{p}], \mathsf{v}^0)\big) \\
(\Gamma_1, \Gamma_2, \dots) \rhd_2 A_= & := & \big(\Gamma_2 \circ (\epsilon, \mathsf{fst}\,\mathsf{v}^0), \mathsf{snd}\,\mathsf{v}^0\big) \\
(\Gamma_1, \Gamma_2, \dots) \rhd_{12} A_= & : & (\Gamma_1, \Gamma_2, \dots) \rhd_1 A_= \circ (\Gamma_1, \Gamma_2, \dots) \rhd_2 A_= \\
 & & \big(\epsilon, (\mathsf{v}^0[\Gamma_1 \circ \mathsf{p}], \mathsf{v}^0)\big) \circ \big(\Gamma_2 \circ (\epsilon, \mathsf{fst}\,\mathsf{v}^0), \mathsf{snd}\,\mathsf{v}^0\big) = \\
 & & \big(\epsilon, (\mathsf{v}^0[\Gamma_1 \circ \Gamma_2 \circ (\epsilon, \mathsf{fst}\,\mathsf{v}^0)], \mathsf{snd}\,\mathsf{v}^0)\big) = \\
 & & \big(\epsilon, (\mathsf{fst}\,\mathsf{v}^0, \mathsf{snd}\,\mathsf{v}^0)\big) = \\
 & & \big(\epsilon, \mathsf{v}^0\big) = \\
 & & (\mathsf{p}, \mathsf{q}) = \\
 & & \mathsf{id} \\
(\Gamma_1, \Gamma_2, \dots) \rhd_{21} A_= & : & (\Gamma_1, \Gamma_2, \dots) \rhd_2 A_= \circ (\Gamma_1, \Gamma_2, \dots) \rhd_1 A_= \\
 & & \big(\Gamma_2 \circ (\epsilon, \mathsf{fst}\,\mathsf{v}^0), \mathsf{snd}\,\mathsf{v}^0\big) \circ \big(\epsilon, (\mathsf{v}^0[\Gamma_1 \circ \mathsf{p}], \mathsf{v}^0)\big) = \\
 & & \big(\Gamma_2 \circ (\epsilon, \mathsf{v}^0[\Gamma_1 \circ \mathsf{p}]), \mathsf{v}^0\big) = \\
 & & \big(\Gamma_2 \circ \Gamma_1 \circ \mathsf{p}, \mathsf{v}^0\big) = \\
 & & (\mathsf{p}, \mathsf{q}) = \\
 & & \mathsf{id} \\
\sigma_= ,_= t_= & : & (\sigma, t) =
\end{array}$$

$$
\begin{aligned}
& && (\Delta_2 \circ (\epsilon, \mathsf{app}\,[\![\sigma]\!]) \circ \Gamma_1, \mathsf{app}\,[\![t]\!][\Gamma_1]) = \\
& && (\Delta_2 \circ (\epsilon, \mathsf{fst}\,\mathsf{v}^0), \mathsf{snd}\,\mathsf{v}^0) \circ (\epsilon, (\mathsf{app}\,[\![\sigma]\!], \mathsf{app}\,[\![t]\!])) \circ \Gamma_1 = \\
& && (\Delta_1, \dots) \rhd_2 A_= \circ (\epsilon, \mathsf{app}\,[\![\sigma, t]\!]) \circ \Gamma_1 \\
& \mathsf{p}_= &:\ & \mathsf{p} = \\
& && \Gamma_2 \circ \Gamma_1 = \\
& && \Gamma_2 \circ (\epsilon, \mathsf{fst}\,\mathsf{v}^0) \circ \big(\epsilon, (\mathsf{v}^0[\Gamma_1 \circ \mathsf{p}], \mathsf{v}^0)\big) = \\
& && \Gamma_2 \circ (\epsilon, \mathsf{app}\,[\![\mathsf{p}]\!]) \circ (\Gamma_1, \dots) \rhd_1 A_= \\
& \mathsf{q}_= &:\ & \mathsf{q} = \mathsf{v}^0 = \\
& && \mathsf{lam}(\mathsf{snd}\,\mathsf{v}^0) = \\
& && (\mathsf{snd}\,\mathsf{v}^0)[\epsilon, (\mathsf{v}^0[\Gamma_1 \circ \mathsf{p}], \mathsf{v}^0)] = \\
& && \mathsf{app}\,[\![\mathsf{q}]\!][(\Gamma_1, \dots) \rhd_1 A_=] \\
& \Pi_= A_= B_= &:\ & \Pi\, A\, B = \\
& && \Pi\, \underline{\mathsf{app}\,[\![A]\!][\Gamma_1]}\ \underline{\mathsf{app}\,[\![B]\!][(\Gamma_1, \dots) \rhd_1 A_=]} = \\
& && \Pi\, \underline{\mathsf{app}\,[\![A]\!][\Gamma_1]}\ \underline{\mathsf{app}\,[\![B]\!][\epsilon, (\mathsf{v}^1, \mathsf{v}^0)][\Gamma_1{}^{\uparrow}]} = \\
& && \underline{\mathsf{app}\,[\![\Pi\, A\, B]\!]}[\Gamma_1] \\
& \mathsf{lam}_= t_= &:\ & \mathsf{lam}\, t = \\
& && \mathsf{lam}\,(\mathsf{app}\,[\![t]\!][(\Gamma_1, \dots) \rhd_1 A_=]) = \\
& && \mathsf{lam}\,(\mathsf{app}\,[\![t]\!][\epsilon, (\mathsf{v}^1, \mathsf{v}^0)][\Gamma_1{}^{\uparrow}]) = \\
& && \mathsf{lam}\,(\mathsf{app}\,[\![t]\!][\epsilon, (\mathsf{v}^1, \mathsf{v}^0)])[\Gamma_1] = \\
& && \mathsf{app}\,(\mathsf{lam}\,(\mathsf{lam}\,([\![t]\!][\epsilon]\,\$(\mathsf{v}^1, \mathsf{v}^0))))[\Gamma_1] = \\
& && \mathsf{app}\,[\![\mathsf{lam}\, t]\!][\Gamma_1] \\
& \mathsf{app}_= t_= &:\ & \mathsf{app}\, t = \\
& && \mathsf{app}\,(\mathsf{app}\,[\![t]\!][\Gamma_1]) = \\
& && \mathsf{app}\,(\mathsf{app}\,[\![t]\!])[\Gamma_1{}^{\uparrow}] = \\
& && \mathsf{app}\,(\mathsf{app}\,[\![t]\!])[\epsilon, \mathsf{v}^1, \mathsf{v}^0][\Gamma_1{}^{\uparrow}] = \\
& && \mathsf{app}\,(\mathsf{app}\,[\![t]\!])[\epsilon, \mathsf{v}^0[\Gamma_1 \circ \mathsf{p}], \mathsf{v}^0] = \\
& && \mathsf{app}\,(\mathsf{app}\,[\![t]\!])[\epsilon, \mathsf{fst}\,\mathsf{v}^0, \mathsf{snd}\,\mathsf{v}^0][\epsilon, (\mathsf{v}^0[\Gamma_1 \circ \mathsf{p}], \mathsf{v}^0)] = \\
& && \mathsf{app}\,(\mathsf{app}\,[\![t]\!])[\epsilon, \mathsf{fst}\,\mathsf{v}^1, \mathsf{v}^0][\mathsf{id}, \mathsf{snd}\,\mathsf{v}^0][\epsilon, (\mathsf{v}^0[\Gamma_1 \circ \mathsf{p}], \mathsf{v}^0)] = \\
& && \mathsf{app}\,(\mathsf{app}\,[\![t]\!][\epsilon, \mathsf{fst}\,\mathsf{v}^0])[\mathsf{id}, \mathsf{snd}\,\mathsf{v}^0][(\Gamma_1, \dots) \rhd_1 A_=] = \\
& && ([\![t]\!][\epsilon]\,\$\,\mathsf{fst}\,\mathsf{v}^0\,\$\,\mathsf{snd}\,\mathsf{v}^0)[(\Gamma_1, \dots) \rhd_1 A_=] = \\
& && \mathsf{app}\,(\mathsf{lam}\,([\![t]\!][\epsilon]\,\$\,\mathsf{fst}\,\mathsf{v}^0\,\$\,\mathsf{snd}\,\mathsf{v}^0))[(\Gamma_1, \dots) \rhd_1 A_=] = \\
& && \mathsf{app}\,[\![\mathsf{app}\, t]\!][(\Gamma_1, \dots) \rhd_1 A_=] \\
& \Sigma_= A_= B_= &:\ & \Sigma\, A\, B = \\
& && \Sigma\, \underline{\mathsf{app}\,[\![A]\!][\Gamma_1]}\ \underline{\mathsf{app}\,[\![B]\!][(\Gamma_1, \dots) \rhd_1 A_=]} = \\
& && \Sigma\, \underline{\mathsf{app}\,[\![A]\!][\Gamma_1]}\ \underline{\mathsf{app}\,[\![B]\!][\epsilon, (\mathsf{v}^1, \mathsf{v}^0)][\Gamma_1{}^{\uparrow}]} =
\end{aligned}
$$

$$
\begin{array}{lll}
 & & \underline{\mathsf{app}\,[\![\Sigma\,A\,B]\!]}[\Gamma_1] \\
u_{=} ,_{=} v_{=} & : & (u, v) = \\
 & & (\mathsf{app}\,[\![u]\!][\Gamma_1], \mathsf{app}\,[\![v]\!][\Gamma_1]) = \\
 & & (\mathsf{app}\,[\![u]\!], \mathsf{app}\,[\![v]\!])[\Gamma_1] = \\
 & & \mathsf{app}\,[\![u, v]\!][\Gamma_1] \\
\mathsf{fst}_{=}\,t_{=} & : & \mathsf{fst}\,t = \\
 & & \mathsf{fst}\,(\mathsf{app}\,[\![t]\!][\Gamma_1]) = \\
 & & (\mathsf{fst}\,(\mathsf{app}\,[\![t]\!]))[\Gamma_1] = \\
 & & \mathsf{app}\,[\![\mathsf{fst}\,t]\!][\Gamma_1] \\
\mathsf{snd}_{=}\,t_{=} & : & \mathsf{snd}\,t = \\
 & & \mathsf{snd}\,(\mathsf{app}\,[\![t]\!][\Gamma_1]) = \\
 & & (\mathsf{snd}\,(\mathsf{app}\,[\![t]\!]))[\Gamma_1] = \\
 & & \mathsf{app}\,[\![\mathsf{snd}\,t]\!][\Gamma_1] \\
\top_{=} & : & \top = \top[\Gamma_1] = \underline{\mathsf{app}\,(\mathsf{lam}\,(\mathsf{c}\,\top))[\Gamma_1]} = \underline{\mathsf{app}\,[\![\top]\!][\Gamma_1]} \\
\mathsf{tt}_{=} & : & \mathsf{tt} = \mathsf{tt}[\Gamma_1] = \mathsf{app}\,(\mathsf{lam}\,\mathsf{tt})[\Gamma_1] = \mathsf{app}\,[\![\mathsf{tt}]\!][\Gamma_1] \\
\mathsf{U}_{=} & : & \mathsf{U}\,i = \mathsf{U}\,i[\Gamma_1] = \underline{\mathsf{app}\,(\mathsf{lam}\,(\mathsf{c}\,(\mathsf{U}\,i)))}[\Gamma_1] = \underline{\mathsf{app}\,[\![\mathsf{U}\,i]\!]}[\Gamma_1] \\
\underline{a}_{=} & : & \underline{a} = \underline{\mathsf{app}\,[\![a]\!][\Gamma_1]} = \underline{\mathsf{app}\,[\![\underline{a}]\!]}[\Gamma_1] \\
\mathsf{c}_{=}\,A_{=} & : & A = \mathsf{app}\,[\![A]\!][\Gamma_1] = \mathsf{app}\,[\![\mathsf{c}\,A]\!][\Gamma_1] \\
\mathsf{Bool}_{=} & : & \mathsf{Bool} = \underline{\mathsf{c}\,\mathsf{Bool}}[\Gamma_1] = \underline{\mathsf{app}\,[\![\mathsf{Bool}]\!]}[\Gamma_1] \\
\mathsf{true}_{=} & : & \mathsf{true} = \mathsf{true}[\Gamma_1] = \mathsf{app}\,[\![\mathsf{true}]\!][\Gamma_1] \\
\mathsf{false}_{=} & : & \mathsf{false} = \mathsf{false}[\Gamma_1] = \mathsf{app}\,[\![\mathsf{false}]\!][\Gamma_1] \\
\mathsf{if}_{=}\,C_{=}\,u_{=}\,v_{=}\,t_{=} & : & \mathsf{if}\,C\,u\,v\,t = \\
 & & \mathsf{if}\,\underline{\mathsf{app}\,[\![C]\!][(\Gamma \rhd \mathsf{Bool})_1]}\,(\mathsf{app}\,[\![u]\!][\Gamma_1])\,(\mathsf{app}\,[\![v]\!][\Gamma_1]) \\
 & & \quad (\mathsf{app}\,[\![t]\!][\Gamma_1]) = \\
 & & \mathsf{if}\,\underline{\mathsf{app}\,[\![C]\!][\epsilon, (\mathsf{v}^1, \mathsf{v}^0)][{\Gamma_1}^{\uparrow}]}\,(\mathsf{app}\,[\![u]\!][\Gamma_1])\,(\mathsf{app}\,[\![v]\!][\Gamma_1]) \\
 & & \quad (\mathsf{app}\,[\![t]\!][\Gamma_1]) = \\
 & & \mathsf{if}\,\underline{[\![C]\!][\epsilon]\,\$(\mathsf{v}^1, \mathsf{v}^0)}\,(\mathsf{app}\,[\![u]\!])\,(\mathsf{app}\,[\![v]\!])\,(\mathsf{app}\,[\![t]\!])[\Gamma_1] = \\
 & & \mathsf{app}\,[\![\mathsf{if}\,C\,u\,v\,t]\!][\Gamma_1] \\
\mathsf{Id}_{=}\,A_{=}\,u_{=}\,v_{=} & : & \mathsf{Id}\,A\,u\,v = \\
 & & \mathsf{Id}\,\underline{\mathsf{app}\,[\![A]\!][\Gamma_1]}\,(\mathsf{app}\,[\![u]\!][\Gamma_1])\,(\mathsf{app}\,[\![v]\!][\Gamma_1]) \\
 & & \big(\mathsf{Id}\,\underline{\mathsf{app}\,[\![A]\!]}\,(\mathsf{app}\,[\![u]\!])\,(\mathsf{app}\,[\![v]\!])\big)\,[\Gamma_1] \\
 & & \underline{\mathsf{app}\,[\![\mathsf{Id}\,A\,u\,v]\!]}[\Gamma_1] \\
\mathsf{refl}_{=}\,u_{=} & : & \mathsf{refl}\,u = \\
 & & \mathsf{refl}\,(\mathsf{app}\,[\![u]\!][\Gamma_1]) = \\
 & & \mathsf{refl}\,(\mathsf{app}\,[\![u]\!])[\Gamma_1] =
\end{array}
$$

$$
\begin{aligned}
& && \mathsf{app}\,[\![\mathsf{refl}\,u]\!][\Gamma_1] \\
\mathsf{J}_{=}\;C_{=}\,w_{=}\,e_{=} &:&& \mathsf{J}\,C\,w\,e = \\
& && \mathsf{J}\,\underline{\mathsf{app}\,[\![C]\!][(\Gamma \triangleright A \triangleright \ldots)_1]}\,(\mathsf{app}\,[\![w]\!][\Gamma_1])\,(\mathsf{app}\,[\![e]\!][\Gamma_1]) = \\
& && \mathsf{J}\,\underline{\mathsf{app}\,[\![C]\!][\epsilon,(\mathsf{v}^2,\mathsf{v}^1,\mathsf{v}^0)][{\Gamma_1}^{\uparrow\uparrow}]}\,(\mathsf{app}\,[\![w]\!][\Gamma_1])\,(\mathsf{app}\,[\![e]\!][\Gamma_1]) = \\
& && \big(\mathsf{J}\,\underline{\mathsf{app}\,[\![C]\!][\epsilon,(\mathsf{v}^2,\mathsf{v}^1,\mathsf{v}^0)]}\,(\mathsf{app}\,[\![w]\!])\,(\mathsf{app}\,[\![e]\!])\big)\,[\Gamma_1] = \\
& && \mathsf{app}\,[\![\mathsf{J}\,C\,w\,e]\!][\Gamma_1]
\end{aligned}
$$